\documentclass{aa}
\usepackage{graphicx}
\usepackage{txfonts}
\usepackage{natbib}
\bibpunct{(}{)}{;}{a}{}{,}
\usepackage{longtable}
\usepackage{lscape}
\usepackage{times}
\usepackage{textcomp}

\begin{document}

\title{A Sino-German $\lambda$6\ cm polarization survey of the
  Galactic plane}
\subtitle{VI. Discovery of supernova remnants G178.2$-$4.2 and
  G25.1$-$2.3}
\author{X. Y.~Gao\inst{1}, X. H.~Sun\inst{1,2}, J. L.~Han\inst{1},
  W.~Reich\inst{2}, P.~Reich\inst{2}, R.~Wielebinski\inst{2}}

\titlerunning{The $\lambda$6\ cm survey: discovery of two supernova
  remnants}
\authorrunning{X. Y.~Gao et al.}

\offprints{hjl@nao.cas.cn (HJL); bearwards@gmail.com (XYG)}

\institute{ National Astronomical Observatories, CAS, Jia-20 Datun
  Road, Chaoyang District, Beijing 100012, PR China \and
  Max-Planck-Institut f\"{u}r Radioastronomie, Auf dem H\"{u}gel 69,
  53121 Bonn, Germany }

\date{Received; accepted}

\abstract
%context
{Supernova remnants (SNRs) were often discovered in radio surveys of
  the Galactic plane. Because of the surface-brightness limit of
  previous surveys, more faint or confused SNRs await discovery. The
  Sino-German $\lambda$6\ cm Galactic plane survey is a sensitive
  survey with the potential to detect new low surface-brightness
  SNRs.}
%aims
{We want to identify new SNRs from the $\lambda$6\ cm survey map of
  the Galactic plane.}
%Methods
{We searched for new shell-like objects in the $\lambda$6\ cm survey
  maps, and studied their radio emission, polarization, and spectra
  using the $\lambda$6\ cm maps together with the $\lambda$11\ cm and
  $\lambda$21\ cm Effelsberg observations. Extended polarized objects
  with non-thermal spectra were identified as SNRs.}
%Results
{We have discovered two new, large, faint SNRs, G178.2$-$4.2 and
  G25.1$-$2.3, both of which show shell structure. G178.2$-$4.2 has a
  size of $72\arcmin \times 62\arcmin$ with strongly polarized
  emission being detected along its northern shell.  The spectrum of
  G178.2$-$4.2 is non-thermal, with an integrated spectral index of
  $\alpha = -0.48\pm0.13$. Its surface brightness is $\Sigma_{\rm
    1~GHz} = 7.2\times10^{-23}{\rm Wm^{-2} Hz^{-1} sr^{-1}}$, which
  makes G178.2$-$4.2 the second faintest known Galactic
  SNR. G25.1$-$2.3 is revealed by its strong southern shell which has
  a size of $80\arcmin \times 30\arcmin$. It has a non-thermal radio
  spectrum with a spectral index of $\alpha = -0.49\pm0.13$.}
% Conclusions
{Two new large shell-type SNRs have been detected at $\lambda$6\ cm in
  an area of 2200~deg$^{2}$ along the the Galactic plane. This
  demonstrates that more large and faint SNRs exist, but are very
  difficult to detect.}


\maketitle
\section{Introduction}

Supernova remnants (SNRs) are the post-explosion relics of massive
stars that have reached the end of their evolutionary life times.  It
has been predicted that the total number of Galactic SNRs is between
1\,000 and 10\,000 \citep{Berkhuijsen84, Li91, Tammann94}.  However,
up to now just 274 SNRs have been identified \citep{Green09}. Among
these, G192.8$-$1.1 has recently been disapproved as being a SNR
\citep{Gao11x}. This low SNR detection rate results from the
insufficient sensitivity and resolution of available observations.
Diffuse radio emission and discrete source complexes are not uniformly
distributed along the Galactic disk, whose own emission confuses that
from faint SNRs and hinders their identification.

A large number of SNRs have been identified from radio survey maps
\citep[e.g.][]{Reich88c, Brogan06}. Shell-type SNRs can be recognized
by their morphology, non-thermal spectra, and ordered
polarization. Most known Galactic SNRs have a bright shell, with
intrinsic magnetic fields running along the shell. The radio spectrum
of a SNR is often described by single power law, $S_{\nu} \sim
\nu^{\alpha}$. Here $S_{\nu}$ represents the integrated flux density
of a SNR at the observing frequency $\nu$. In general, SNRs have a
spectral index of $\alpha \sim -0.5$, which is expected for SNRs in
the adiabatic expansion phase with a compression factor of four
\citep[e.g.][]{Reich02}. Young SNRs can have steeper spectra ($\alpha
\sim \rm -0.6\ to -0.8$) and show radial magnetic fields
\citep[e.g.][]{Reich02}.

To discover new SNRs, observations of high sensitivity are needed,
which resolve confusing structures. We have conducted the Sino-German
$\lambda$6\ cm polarization survey of the Galactic plane \citep{Sun07,
  Gao10, Sun11, Xiao11} in the region of Galactic longitude $10\degr
\leq \ell \leq 230\degr$ and latitude $|b| \leq 5\degr$, with an
angular resolution of 9$\farcm$5.  The sensitivity of the Urumqi
survey, if extrapolated from 4.8~GHz to 1~GHz with a typical spectral
index of $\alpha = -0.5$, is on average $\Sigma_{\rm 1~GHz} =
4.9\times10^{-23} {\rm W m^{-2} Hz^{-1} sr^{-1}}$, lower than the
surface brightness of the faintest Galactic SNR known to date, which
is $\Sigma_{\rm 1~GHz} = 5.8\times10^{-23} {\rm W m^{-2} Hz^{-1}
  sr^{-1}}$ for G156.2+5.7 \citep{Reich92, Xu07}. Therefore, it should
be possible to detect new faint SNRs in the $\lambda$6\ cm
survey. Here, we report the discovery of two new SNRs, G178.2$-$4.2
and G25.1$-$2.3, and study their radio properties.

\begin{figure*} %[tbh]
\centering
%6cm \\
\includegraphics[angle=-90, width=0.32\textwidth]{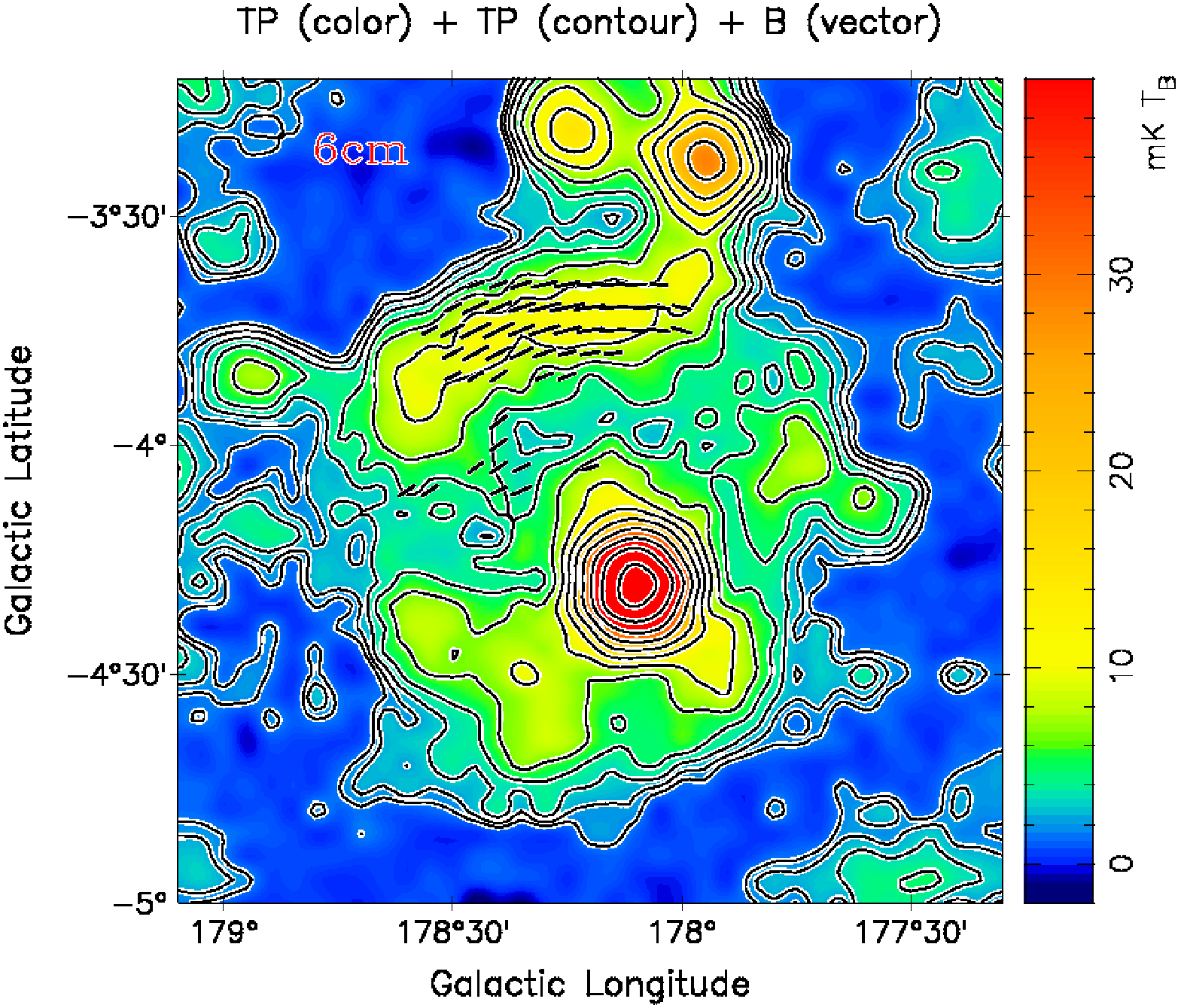}
\includegraphics[angle=-90, width=0.32\textwidth]{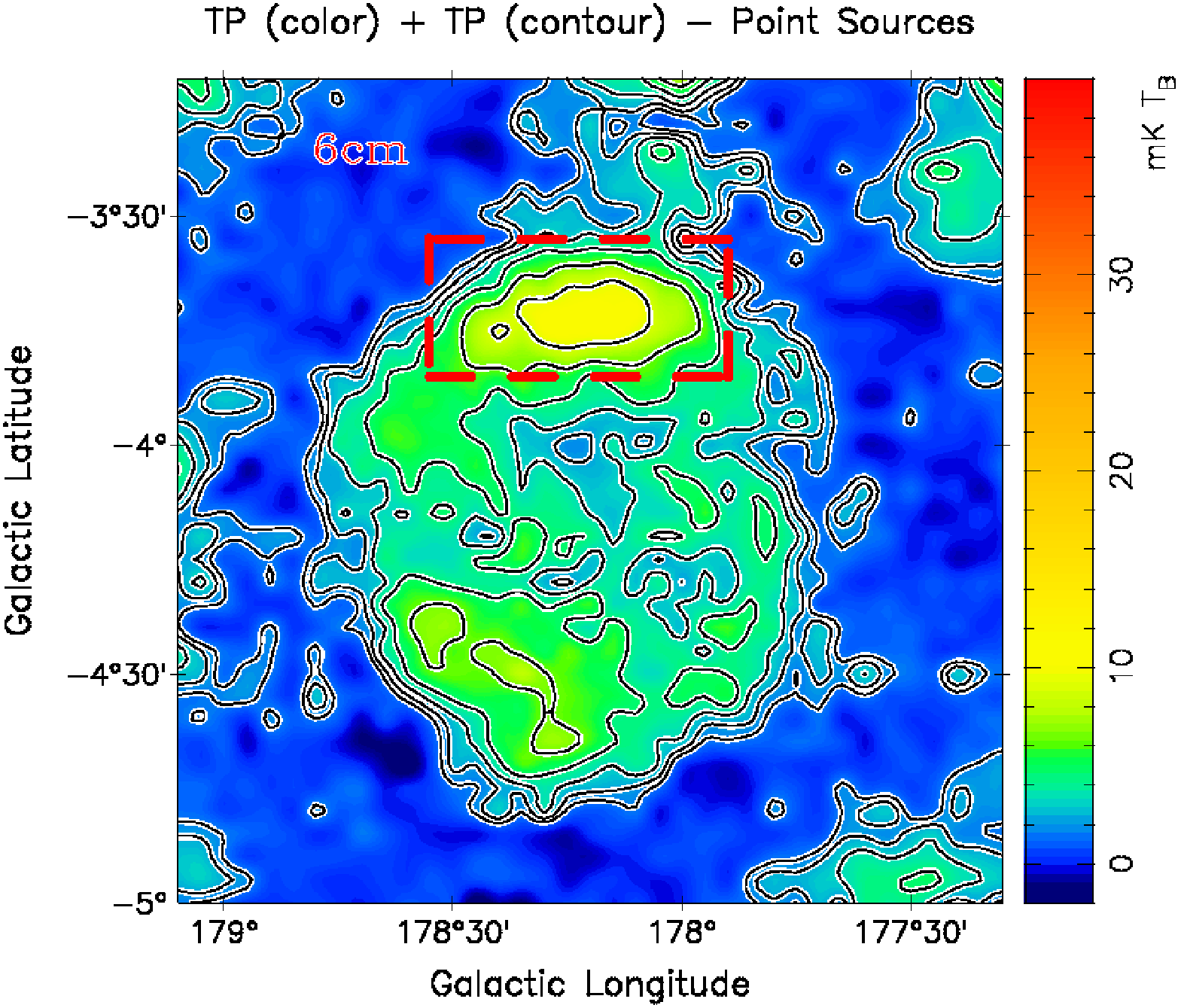}
\includegraphics[angle=-90, width=0.32\textwidth]{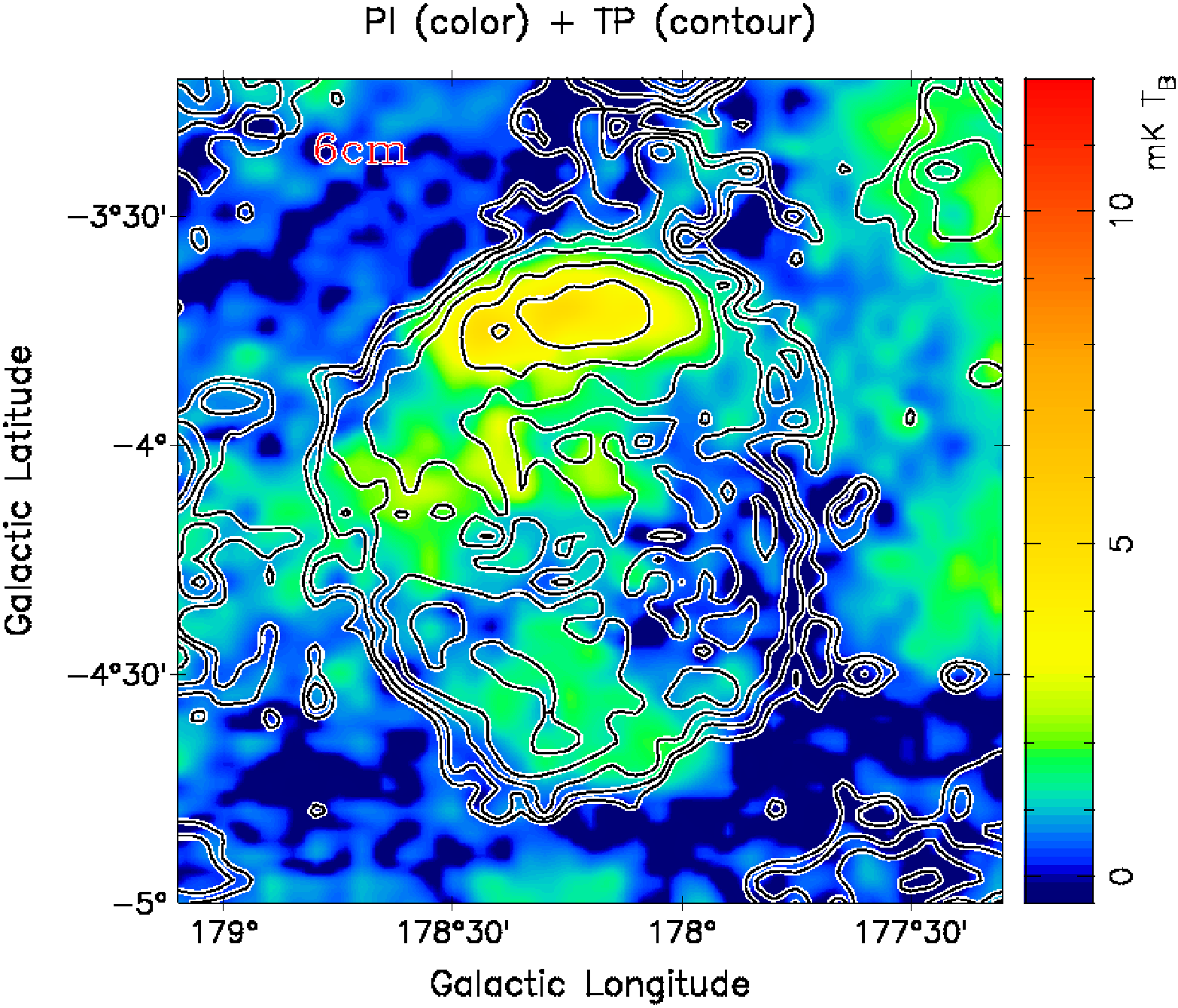}\\[2mm]
% 11cm \\
\includegraphics[angle=-90, width=0.32\textwidth]{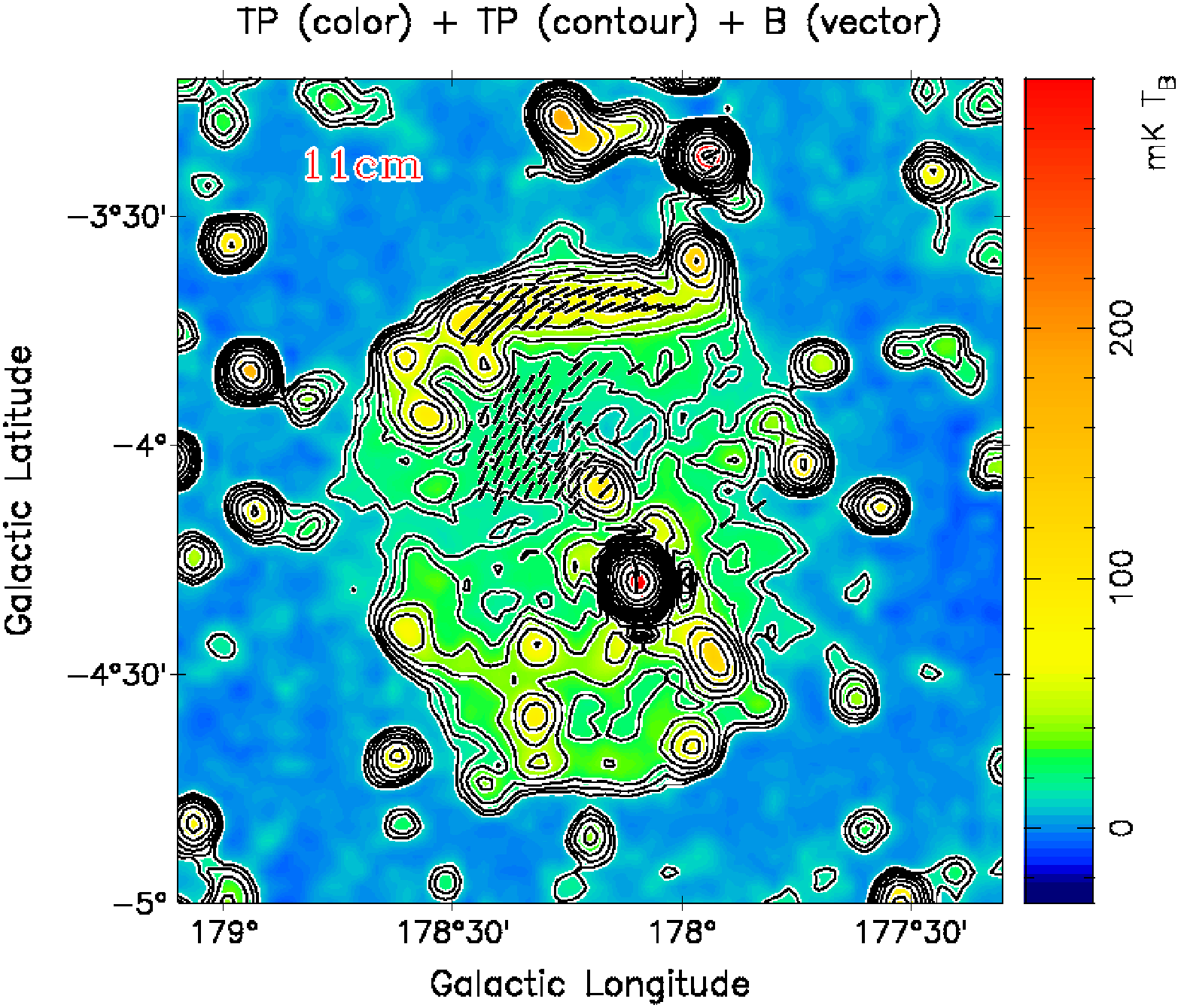}
\includegraphics[angle=-90, width=0.32\textwidth]{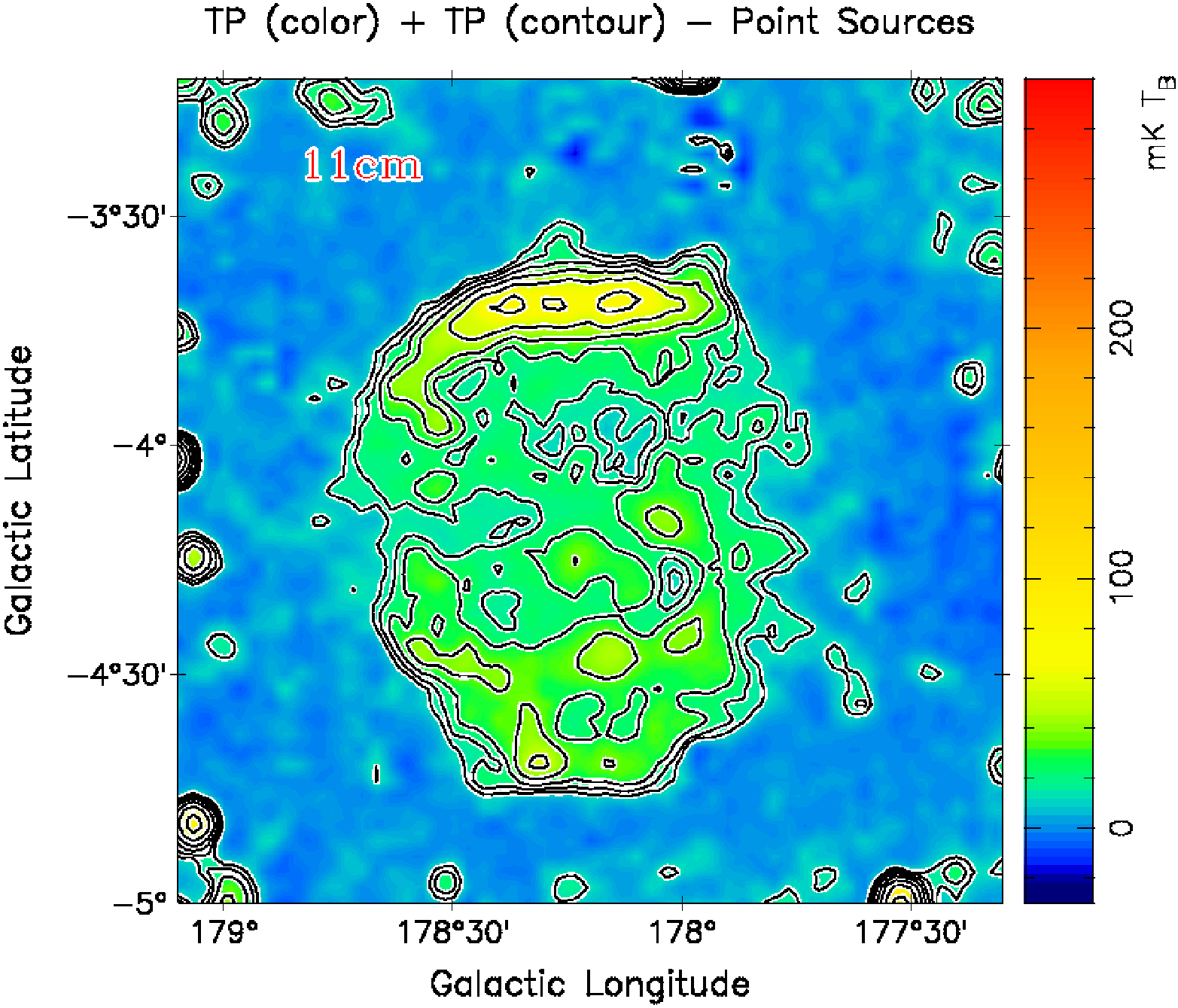}
\includegraphics[angle=-90, width=0.32\textwidth]{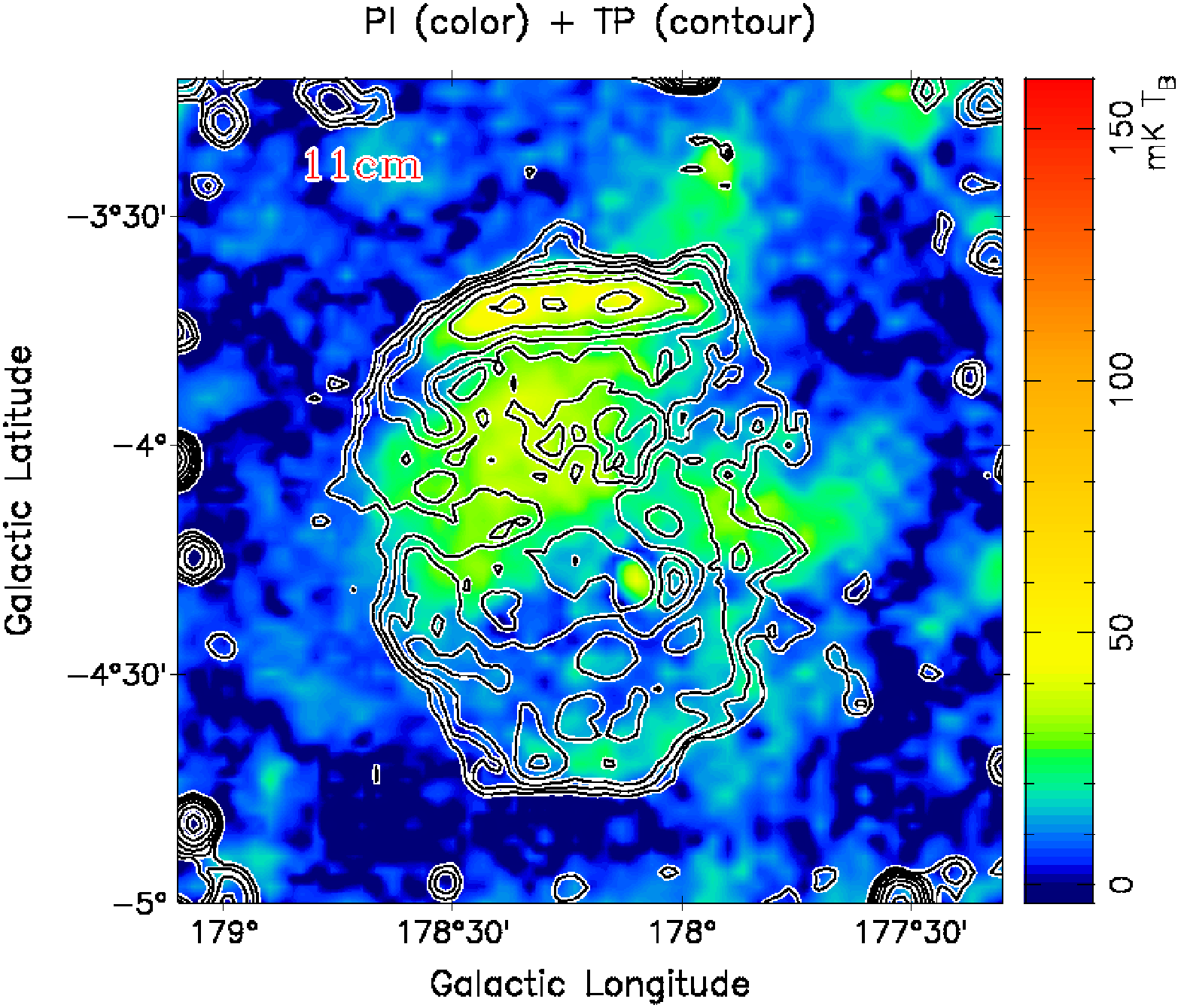}\\[2mm]
% 21cm\\
\includegraphics[angle=-90, width=0.32\textwidth]{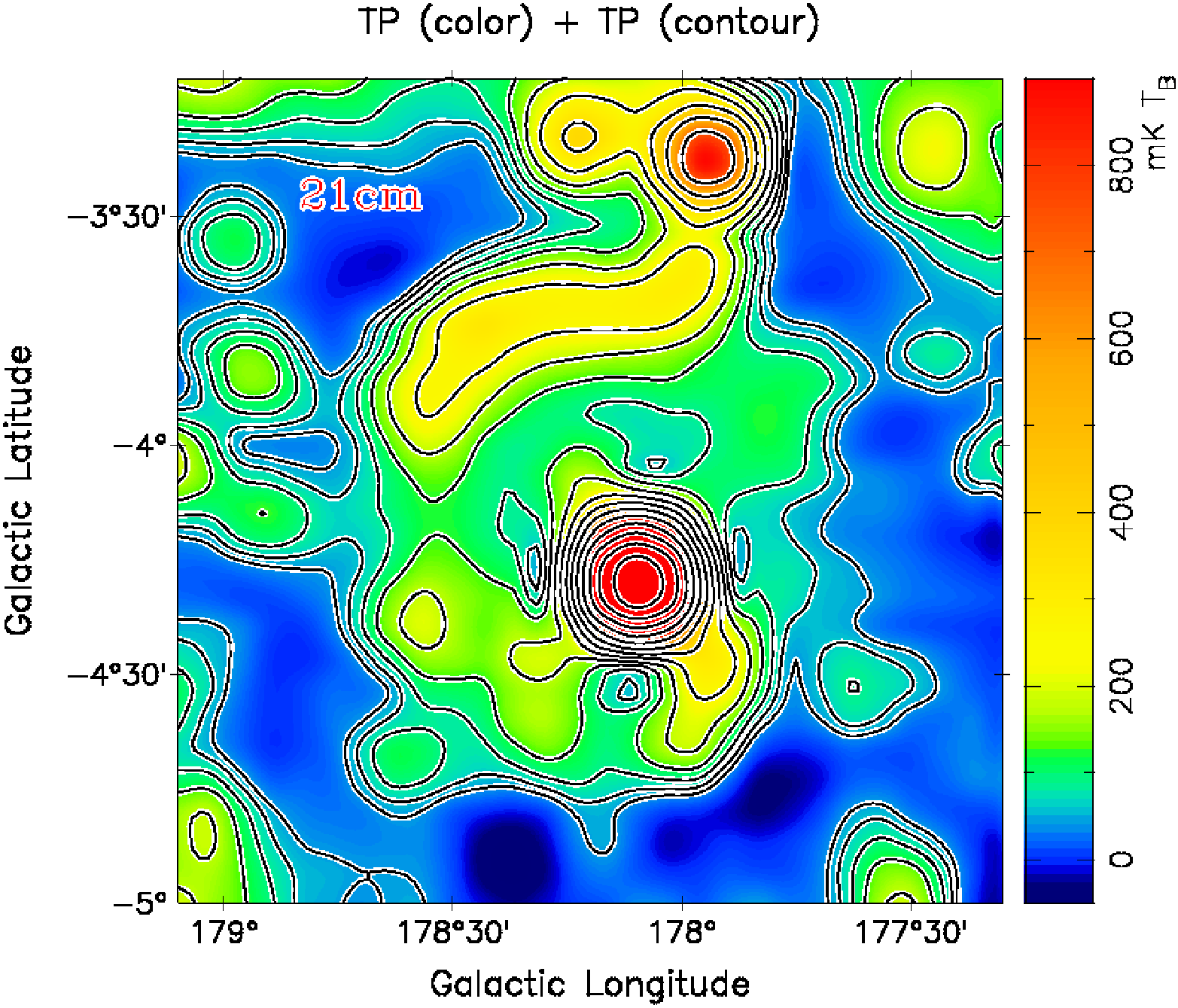}
\includegraphics[angle=-90, width=0.32\textwidth]{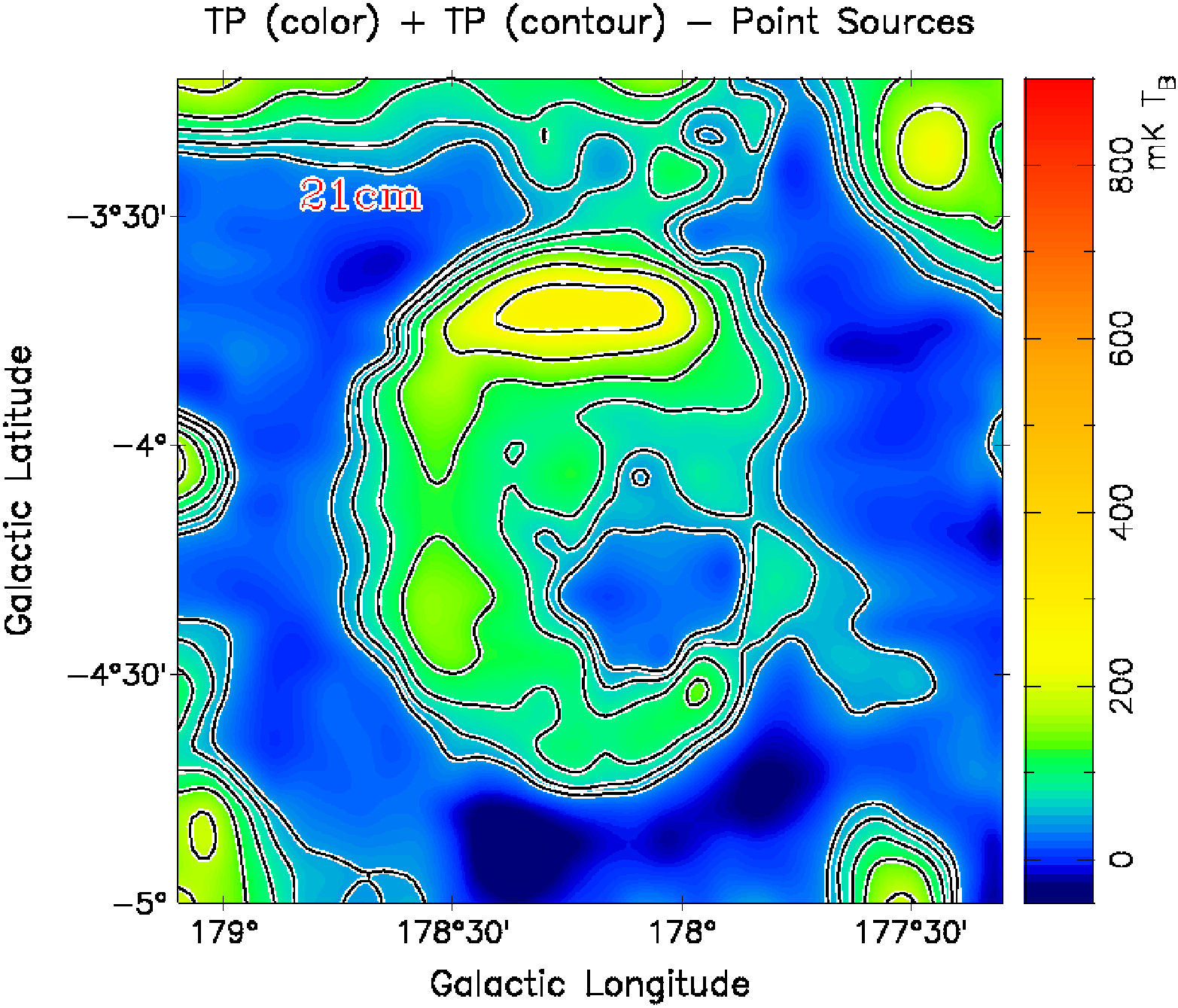}
\includegraphics[angle=-90, width=0.32\textwidth]{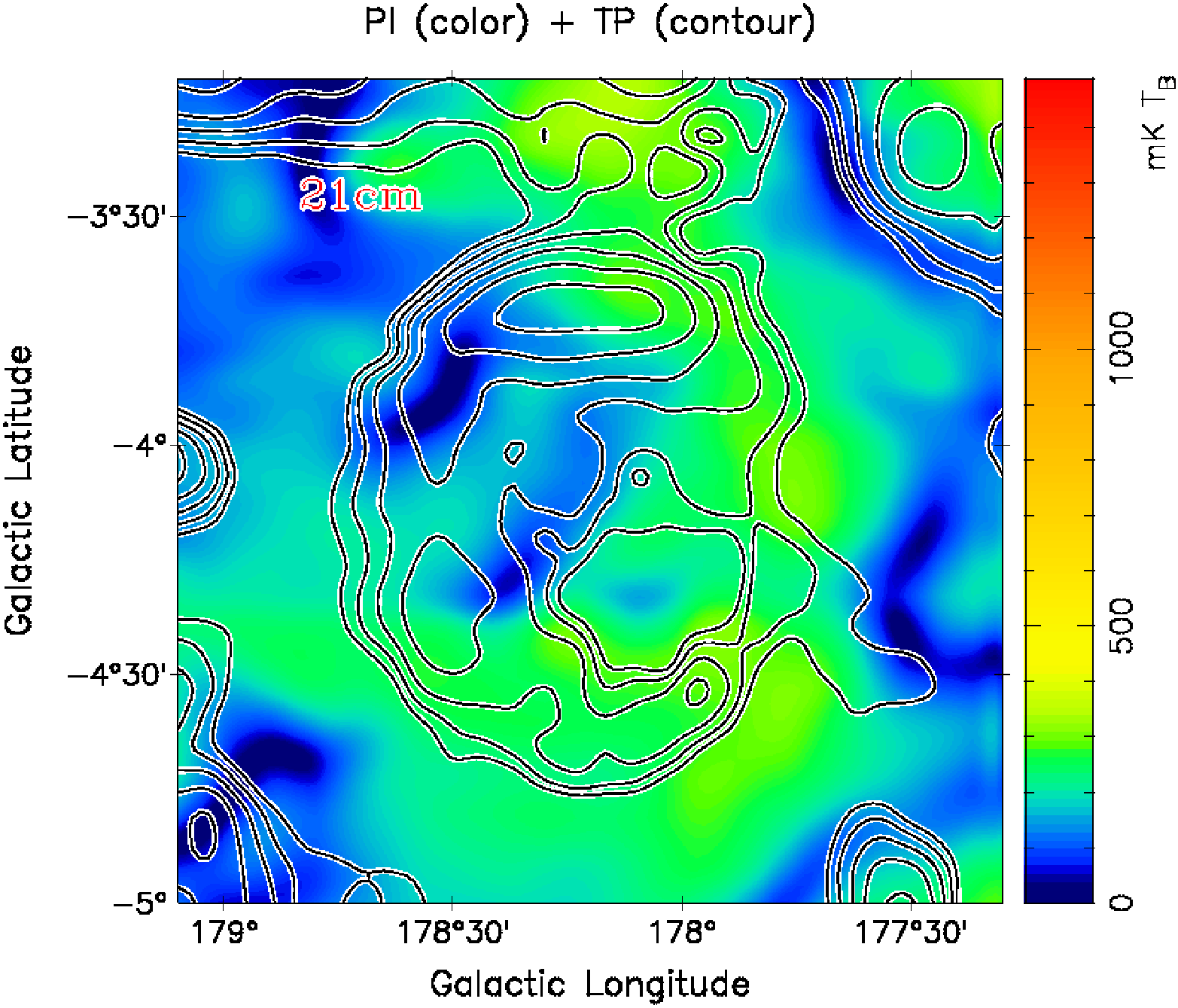}\\
\caption{ Radio images of G178.2$-$4.2 at $\lambda$6\ cm from the
  Sino-German $\lambda$6\ cm polarization survey in the {\it top
    panels}, at $\lambda$11\ cm which we newly observed in the {\it
    middle panels}, and at $\lambda$21\ cm from the Effelsberg Medium
  Latitude Survey (EMLS) \citep{Reich04} in the {\it bottom panels}.
The angular resolutions of the $\lambda$6\ cm, $\lambda$11\ cm and
$\lambda$21\ cm maps are 9.5$\arcmin$, 4.4$\arcmin$ and 9.4$\arcmin$,
respectively.
The {\it left panels} show the total-intensity ($I$) maps in color and
in contours, with observed $B$-vectors overlaid (i.e. the observed
$E$-vectors plus $90\degr$) for polarization intensities $PI >
2.4~{\rm mK}\ T_{B}$ at $\lambda$6\ cm and 32.0~${\rm mK}\ T_{B}$ at
$\lambda$11\ cm. Vectors are not shown at $\lambda$21\ cm. The vector
length is proportional to $PI$. The $I$ contours in the $\lambda$6\ cm
maps are $2^{\frac{n-1}{2}}\times1.6~(4\sigma)~{\rm mK}\ T_{B} $, (n =
1, 2, 3 ...), in the $\lambda$11\ cm maps are
$2^{\frac{n-1}{2}}\times12.8~ (4\sigma)~{\rm mK}\ T_{B}$ (n = 1, 2, 3
...), and in the $\lambda$21\ cm maps are
$2^{\frac{n-1}{2}}\times44.0~(4\sigma)~{\rm mK}\ T_{B} $ (n = 1, 2, 3
...).
The {\it central panels} also display the total-intensity maps, where
strong point-like sources have been subtracted, to show the extended
emission from the SNR more clearly.
The {\it right panels} are the polarization intensity images with the
contours for the total intensity maps with point-like sources
subtracted.
The rectangle in the $\lambda$6\ cm image in the {\it top central
  panel} outlines the area used for the TT-plot spectral analysis
displayed in Fig.~\ref{tt}.}
\label{g178_maps}
\end{figure*}

\begin{figure*}
%\centering
%6cm \\
\includegraphics[angle=-90, width=0.325\textwidth]{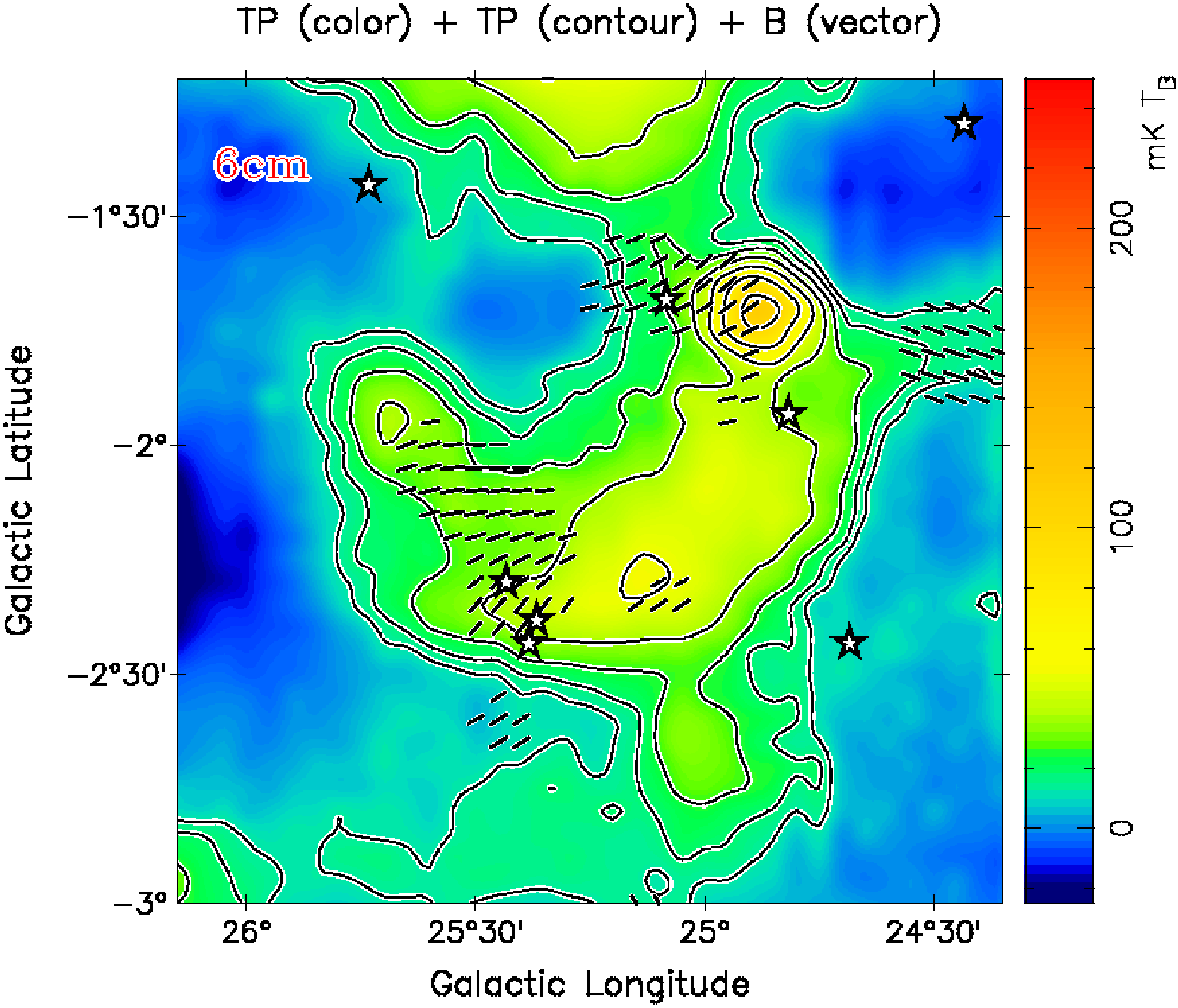}
\includegraphics[angle=-90, width=0.325\textwidth]{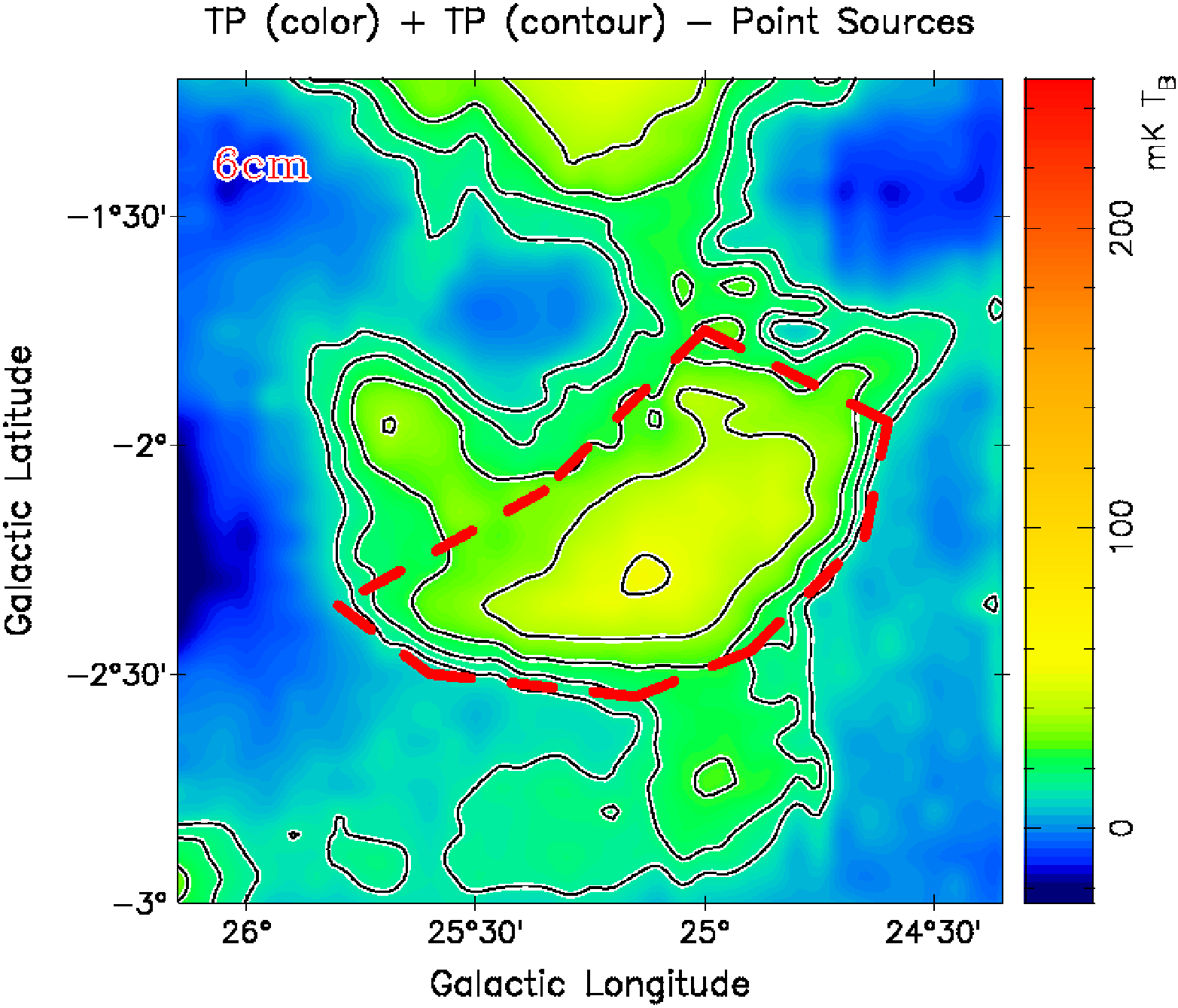}
\includegraphics[angle=-90, width=0.325\textwidth]{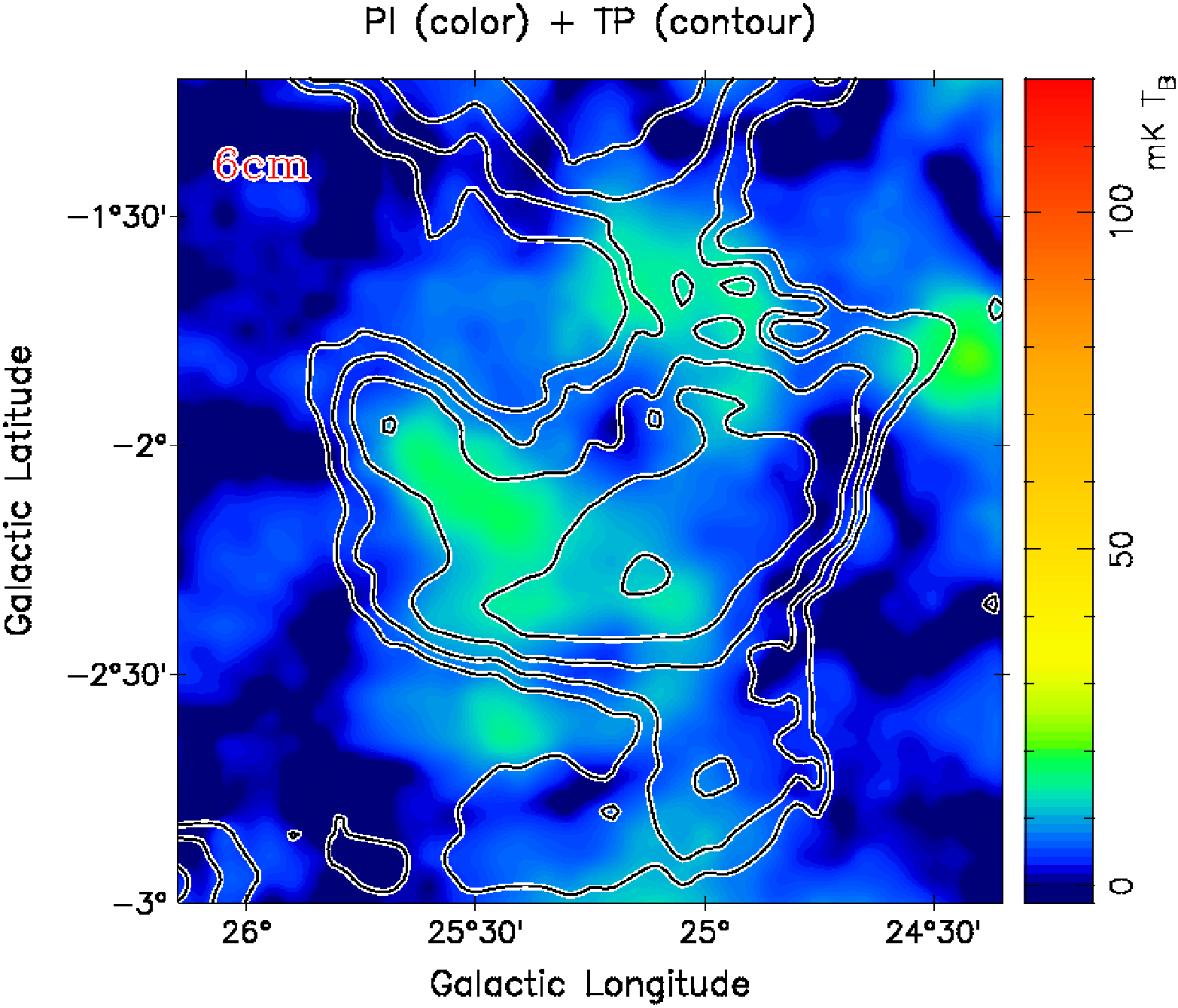}\\[2mm]
% 11cm \\
\includegraphics[angle=-90, width=0.325\textwidth]{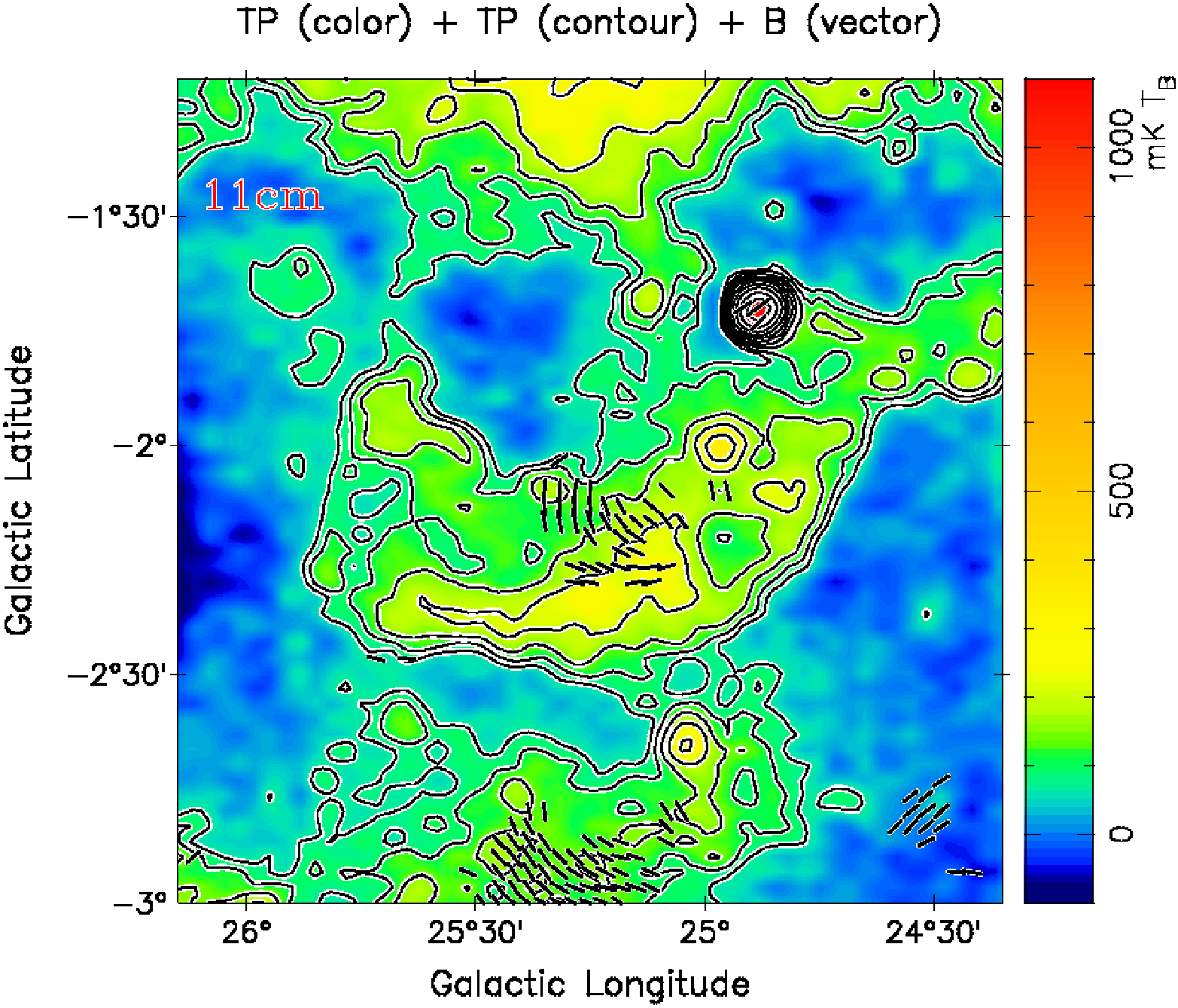}
\includegraphics[angle=-90, width=0.325\textwidth]{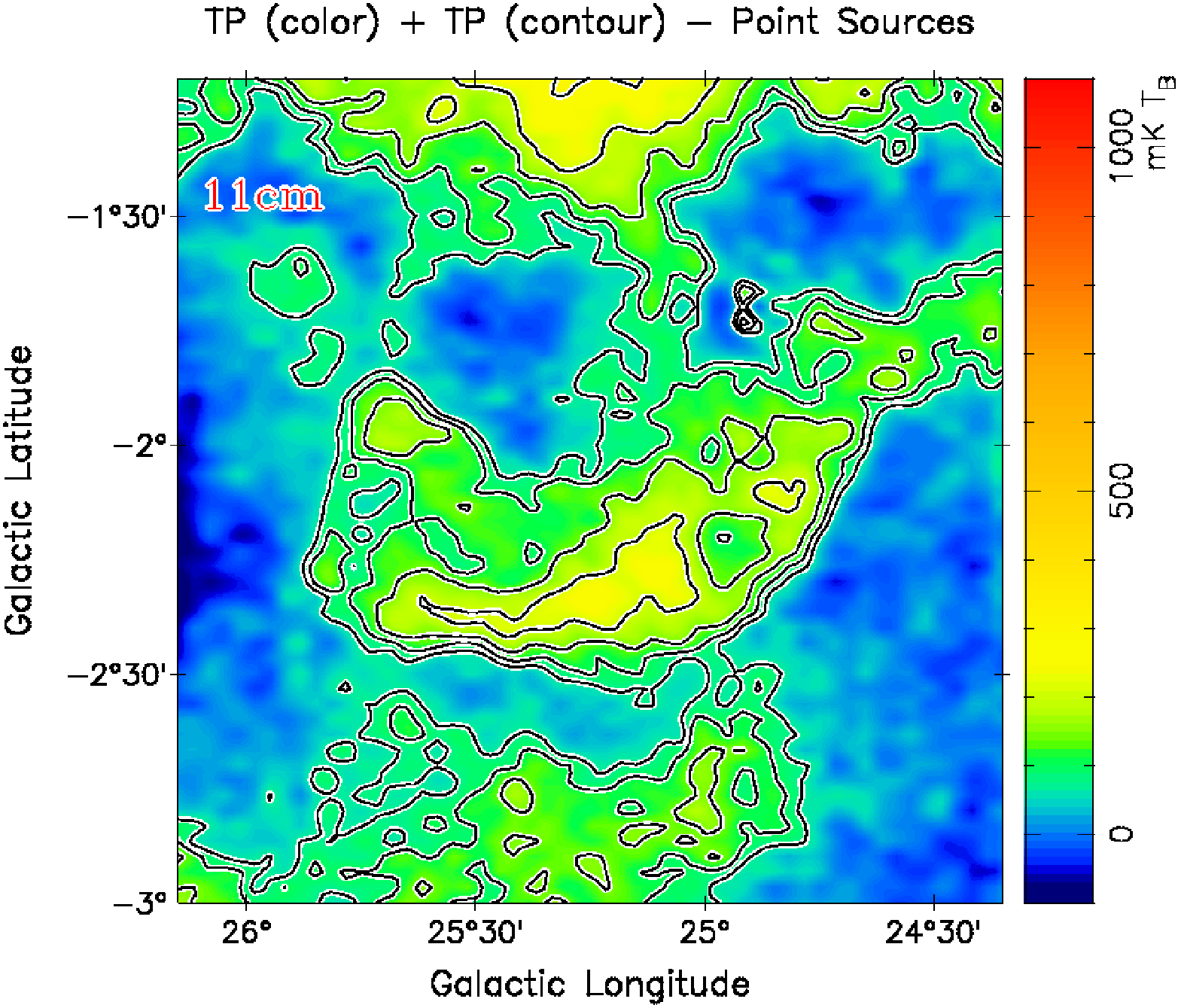}
\includegraphics[angle=-90, width=0.325\textwidth]{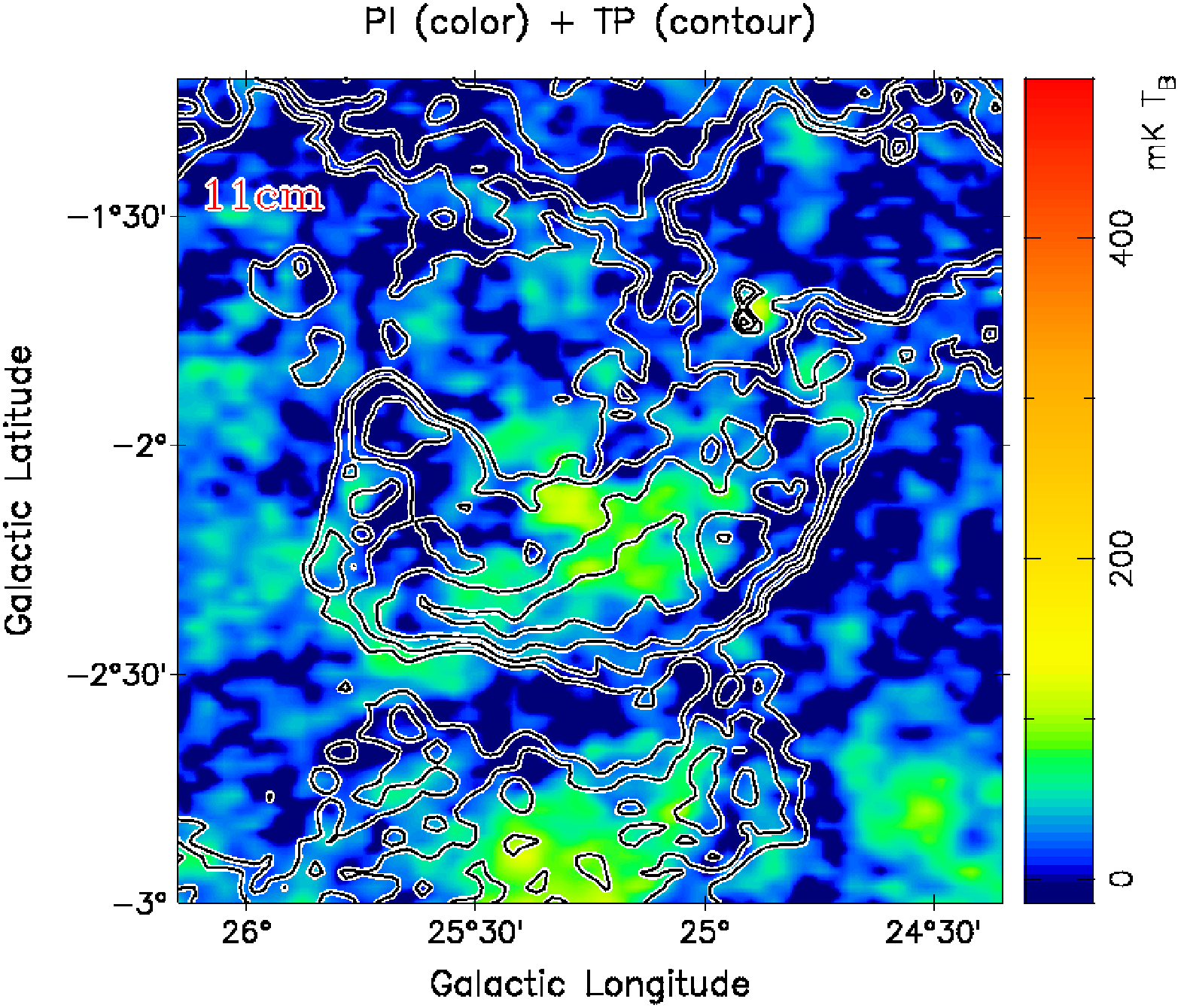}\\[2mm]
% 21cm \\
\includegraphics[angle=-90, width=0.325\textwidth]{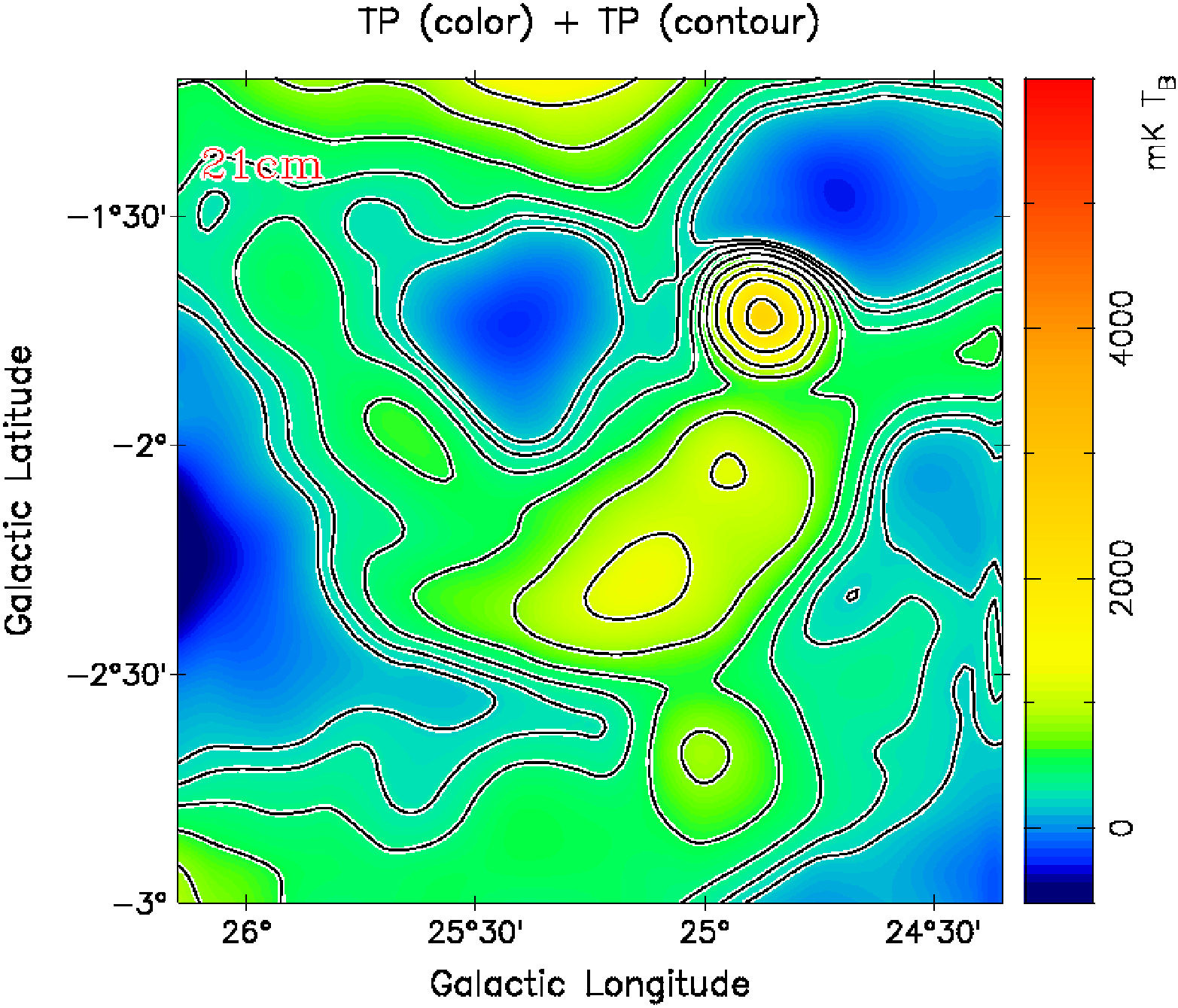}
\includegraphics[angle=-90, width=0.325\textwidth]{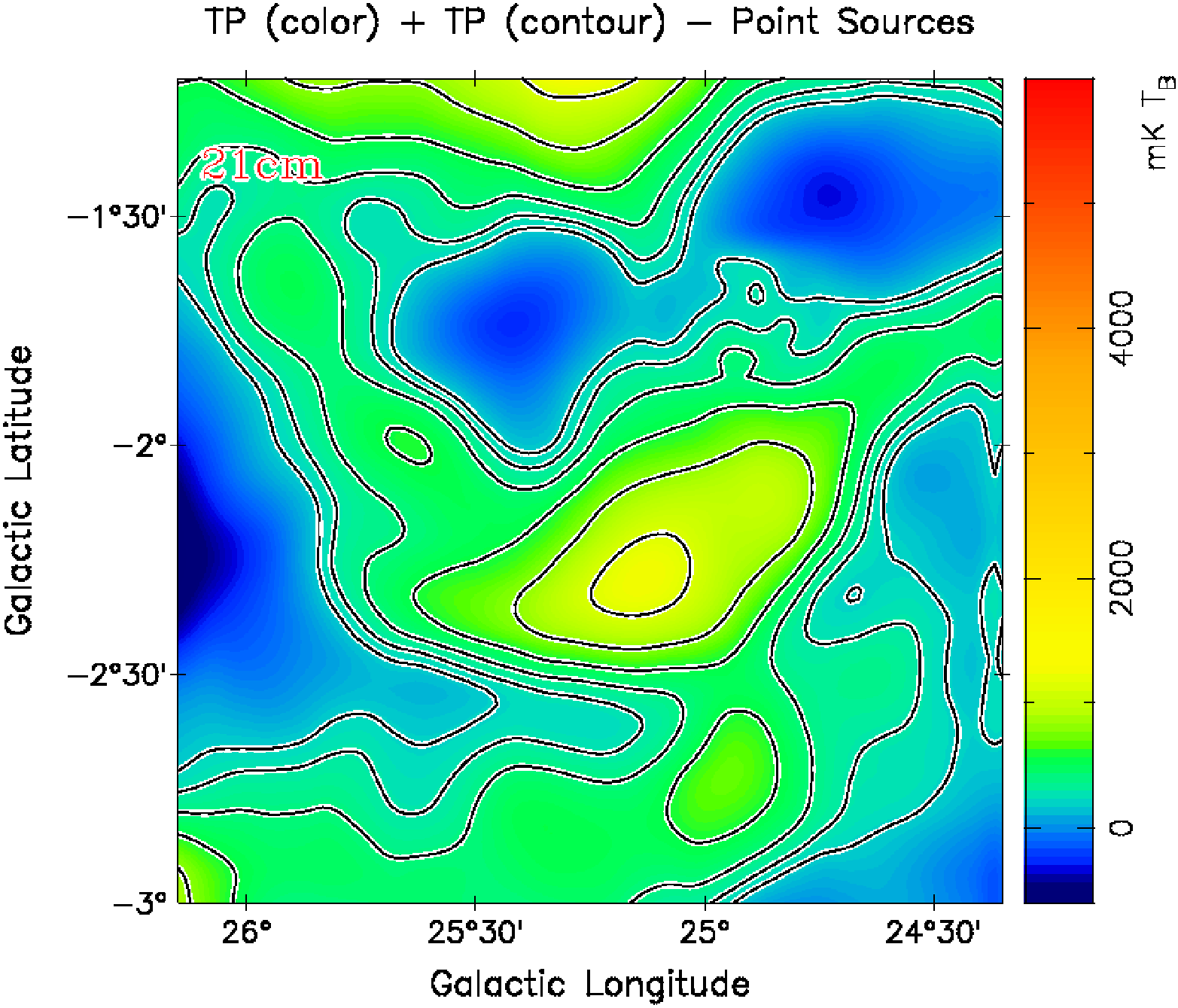}
\caption{Same as Fig.~\ref{g178_maps}, but for G25.1$-$2.3. Here the
  $\lambda$11\ cm and $\lambda$21\ cm maps were extracted from the
  Effelsberg $\lambda$11\ cm survey \citep{Reich9011} and
  $\lambda$21\ cm survey \citep{Reich9021}, respectively. The contours
  in the $\lambda$6\ cm total-intensity maps are $
  2^{\frac{n-1}{2}}\times12.8~(4\sigma)~{\rm mK}\ T_{B} $ (n = 1, 2, 3
  ...), in the $\lambda$11\ cm total-intensity maps are
  $2^{\frac{n-1}{2}}\times72.0~(4\sigma)~{\rm mK}\ T_{B} $ (n =1, 2, 3
  ...), and in the $\lambda$21\ cm total-intensity maps are
  $2^{\frac{n-1}{2}}\times200.0~(4\sigma)~{\rm mK}\ T_{B} $ (n = 1, 2,
  3 ...). The polarization intensity threshold for the $B$-vectors is
  12.0~${\rm mK}\ T_{B}$ at $\lambda$6\ cm and ${\rm 75~mK}\ T_{B}$ at
  $\lambda$11\ cm. No polarization data at $\lambda$21\ cm are
  presently available. The star-symbols in the $\lambda$6\ cm maps
  ({\it top left panel}) indicate known pulsars in the field. The
  polygon in the $\lambda$6\ cm image in {\it top central panel}
  outlines the area for the TT-plot spectral analysis shown in
  Fig.~\ref{tt}.}
\label{g25_maps}
\end{figure*}

\section{The $\lambda$6\ cm maps of the two new SNRs}

The Sino-German $\lambda$6\ cm polarization survey of the Galactic
plane was conducted by using the Urumqi 25-m radio telescope. The
$\lambda$6\ cm maps are not only important for studies of the diffuse
Galactic emission \citep{Sun11, Xiao11}, but also for studies of
Galactic sources like SNRs \citep[e.g.][]{Sun06, Xu07, Shi08a, Xiao08,
  Xiao09, Gao11x}.

We have identified two extended shell-like sources, G178.2$-$4.2 and
G25.1$-$2.3, in the survey maps.  We used the ``background filtering''
technique \citep{Sofue79} with a filter beam size larger than that of
the objects to remove the diffuse large-scale Galactic radio
emission. G178.2$-$4.2 has a size of $72\arcmin \times 62\arcmin$, and
G25.1$-$2.3 has a size of $80\arcmin \times 30\arcmin$. Following
\citet{Gao11x} we measured the mean values at the corners of the
Stokes $I$, $U$, and $Q$ maps in areas without obvious structures. A
hyper plane defined by these corner mean values was subtracted to get
the `intrinsic' total intensity ($I$) and polarization intensity
($PI$) images of the objects (Figs.~\ref{g178_maps} and
\ref{g25_maps}). The $PI$ values were calculated as $PI =
(U^{2} + Q^{2} - 1.2 \sigma_{U,Q}^{2})^{1/2}$ following \citet{Wardle74}. To
study these objects and obtain their integrated flux densities, it is
necessary to first subtract point-like sources in the field of the
objects.  Obvious discrete point-like radio sources were directly
subtracted from the Urumqi $\lambda$6\ cm, and also the Effelsberg
$\lambda$11\ cm and $\lambda$21\ cm maps as we will use in the next
Section, by Gaussian fitting.  Some unresolved sources were identified
in the NVSS catalog \citep{Condon98}. The flux densities of these
sources in our observing bands were extrapolated from the NVSS flux
density at 1.4~GHz, $S_{\rm 1.4GHz}$, using a spectral index either
derived between the flux densities from the NVSS and the Effelsberg
$\lambda$11\ cm survey or quoted from \citet{Vollmer05}. Otherwise,
the mean spectral index for a large sample of radio sources in the
NVSS and WSRT surveys, $\alpha = -0.9$ \citep{Zhang03}, was used, if
the spectral index of a source could not be determined. The
total-intensity maps with point-like sources subtracted are shown in
the central panels of Fig.~\ref{g178_maps} and~\ref{g25_maps}.

\begin{table}
\centering
\caption{Flux densities and the spectral indices of two SNRs.}
\label{tab}
{\begin{tabular}{rrr}\hline\hline 
Objects                                                                           &G178.2$-$4.2                       &G25.1$-$2.3 \\
\hline
$\ell$ ($\degr$)                                                               &178.2                                    &25.1 \\
$b$ ($\degr$)                                                                 &$-$4.2                                   &$-$2.3 \\ 
RA$_{J2000}$ (H M S)                                                   &05 25 06                               &18 45 09 \\
DEC$_{J2000}$ ($\degr$ $\arcmin$ $\arcsec$)             &28 11 02                               &$-$07 59 42 \\
Size (arcmin $\times$ arcmin)                                       &72$\times$62                       &80$\times$30\\
$S_{\rm 6cm}$ (Jy)$\ast$                                              &1.0$\pm$0.1                         &3.7$\pm$0.4 \\
$S_{\rm 11cm}$ (Jy)$\ast$                                            &1.6$\pm$0.2                        &4.7$\pm$0.5 \\
$S_{\rm 21cm}$ (Jy)$\ast$                                            &1.8$\pm$0.2                        &6.7$\pm$0.7 \\
Spectral index $\alpha$                                                  &$-$0.48$\pm$0.13                &$-$0.49$\pm$0.13 \\ 
\hline
\multicolumn{3}{p{.4\textwidth}}{$\ast$: central observing frequencies can be found in the labels of Fig.~\ref{tt}.}
\end{tabular}}
\end{table}

\subsection{G178.2$-$4.2}

G178.2$-$4.2 is located in the anti-center region of the Galaxy and
displays a circular morphology with a prominent shell in its
north. Three strong radio sources, NVSS J052423+281232, NVSS
J052427+281255, and NVSS J052432+281313 are located in the center of
the SNR, which are the three components of the double-sided radio
source 3C 139.2~\citep{Leahy84,Liu02} and not related to the SNR.
Subtracting these and other point-like sources, the extended emission
of the entire SNR G178.2$-$4.2 is clearly visible (see the {\it
  central panels} of Fig.~\ref{g178_maps}). Strong polarized emission
from the northern shell of G178.2$-$4.2 is detected at $\lambda$6\ cm,
with the B-field directions orientated tangential to the shell.

\subsection{G25.1$-$2.3}

A well-pronounced half-shell structure of 80$\arcmin$ in length is
visible at $\ell = 25\fdg1$, $b = -2\fdg3$ after strong diffuse
emission from the Galactic plane is filtered out (see
Fig.~\ref{g25_maps}).  The background radio source centered at $\ell =
24\fdg9, b = -1\fdg7$ is composed by NVSS J184245$-$075613 and NVSS
J184249$-$075604 \citep{Condon98} and is subtracted. It is not clear
whether the extended structure around $\ell = 25\fdg65, b = -1.95$ is
a part of G25.1$-$2.3, because its spectral index from a TT-plot
analysis remains inconclusive. In this paper we only discuss the radio
properties of the shell region outlined in Fig.~\ref{g25_maps}.

\section{Radio properties of G178.2$-$4.2 and G25.1$-$2.3}

The radio morphologies of G178.2$-$4.2 and G25.1$-$2.3 at
$\lambda$6\ cm, as well as the properties discussed below, suggest
that both objects are SNRs.
Using the radio maps at three bands, i.e. $\lambda$6\ cm observed by
using the Urumqi 25-m telescope as described above, $\lambda$11\ cm
and $\lambda$21\ cm by the Effelsberg 100-m telescope, we hereby study
the radio spectrum and polarization properties of G178.2$-$4.2 and
G25.1$-$2.3.

\begin{figure}
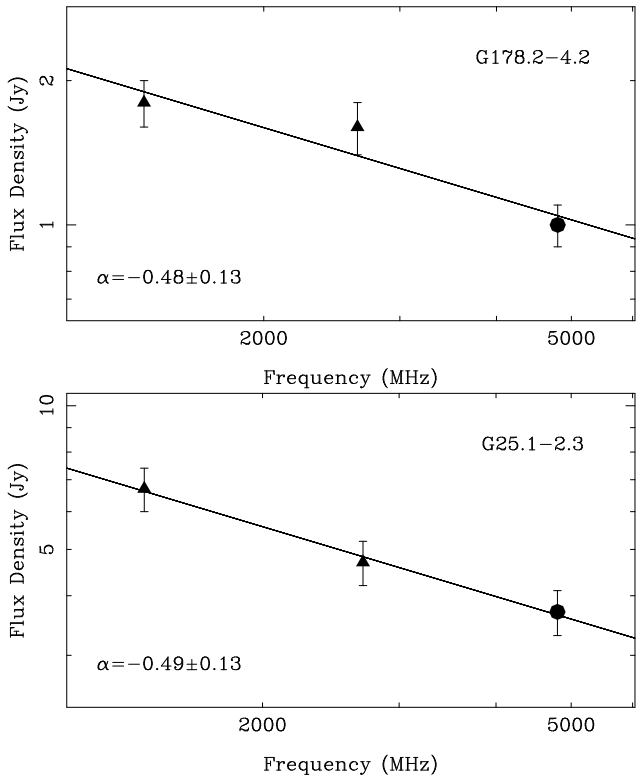

\includegraphics[angle=-90, width=0.45\textwidth]{G178_spectrum.ps}
\includegraphics[angle=-90, width=0.45\textwidth]{G25_spectrum.ps}
\caption{Integrated flux densities and radio spectra of G178.2$-$4.2
  ({\it upper panel}) and the outlined area of G25.1$-$2.3 ({\it lower
    panel}).}
\label{spec}
\end{figure}

We extracted the maps of G178.2$-$4.2 and G25.1$-$2.3 from the
Effelsberg Galactic plane surveys at $\lambda$11\ cm \citep{Fuerst90,
  Reich9011} and $\lambda$21\ cm \citep{Reich9021, Reich97}, see
Fig.~\ref{g178_maps} and Fig.~\ref{g25_maps}. The angular resolution
is 4$\farcm$4 for the $\lambda$11\ cm Effelsberg observations and
9$\farcm$4 for the $\lambda$21\ cm observations. The $\lambda$21\ cm
map of G178.2$-$4.2 was extracted from an unpublished section of the
`Effelsberg Medium Latitude Survey (EMLS)' \citep{Reich04}.

We noticed, however, that G178.2$-$4.2 was poorly traced in the
$\lambda$11\ cm survey map due to its limited sensitivity. The
northern part of G178.2$-$4.2 is clearly detected near the edge of the
recent Effelsberg $\lambda$11\ cm map of S147 \citep{Xiao08} made with
a new receiver. We therefore observed G178.2$-$4.2 again in March 2011
with the Effelsberg 100-m telescope for the new $\lambda$11\ cm map.
The 80~MHz bandwidth of the $\lambda$11\ cm receiver was connected to
an 8-channel polarimeter centered at 2639~MHz. The lowest 10~MHz
channel was corrupted by interference and could not be used. The radio
source, 3C286, was used as the main calibrator assuming $S_{\rm 11cm}
= 10.4$~Jy and 9.9\% linear polarization and a polarization angle of
33$\degr$. We obtained six full coverages of the $2\degr \times
2\degr$ field with in total 4~sec integration time per $2\arcmin$
pixel. We added the maps of seven channels without interference, and
applied the standard data reduction and calibration procedures as
already described by \citet{Xiao08, Xiao09}. The rms-noise measured in
emission-free areas was 3.2~${\rm mK}\ T_{B}$ in the total-intensity
and 2.9~${\rm mK}\ T_{B}$ in the polarization-intensity map. The
G178.2$-$4.2 $\lambda$11\ cm maps in Fig.~\ref{g178_maps} are from the
new observations, rather than extracted from the $\lambda$11\ cm
survey data.

We measured the integrated flux densities of G178.2$-$4.2 and
G25.1$-$2.3 from the radio maps at $\lambda$6\ cm, $\lambda$11\ cm,
and $\lambda$21\ cm. We got $S_{\rm 6cm} = 1.0\pm0.1$~Jy, $S_{\rm
  11cm} = 1.6\pm0.2$~Jy, and $S_{\rm 21cm} = 1.8\pm0.2$~Jy for
G178.2$-$4.2, and $S_{\rm 6cm} = 3.7\pm0.4$~Jy, $S_{\rm 11cm} =
4.7\pm0.5$~Jy, and $S_{\rm 21cm} = 6.7\pm0.7$~Jy for the outlined area
of G25.1$-$2.3 (see Fig.~\ref{g25_maps}). The spectral indices fitted
from these integrated flux densities (see Fig.~\ref{spec}) are $\alpha
= -0.48\pm0.13$ for G178.2$-$4.2 and $\alpha = -0.49\pm0.13$ for the
shell of G25.1$-$2.3, which indicate the non-thermal nature of radio
emission of these two objects. These results are summarized in
Table~\ref{tab}.

\begin{figure}
\includegraphics[angle=-90, width=0.45\textwidth]{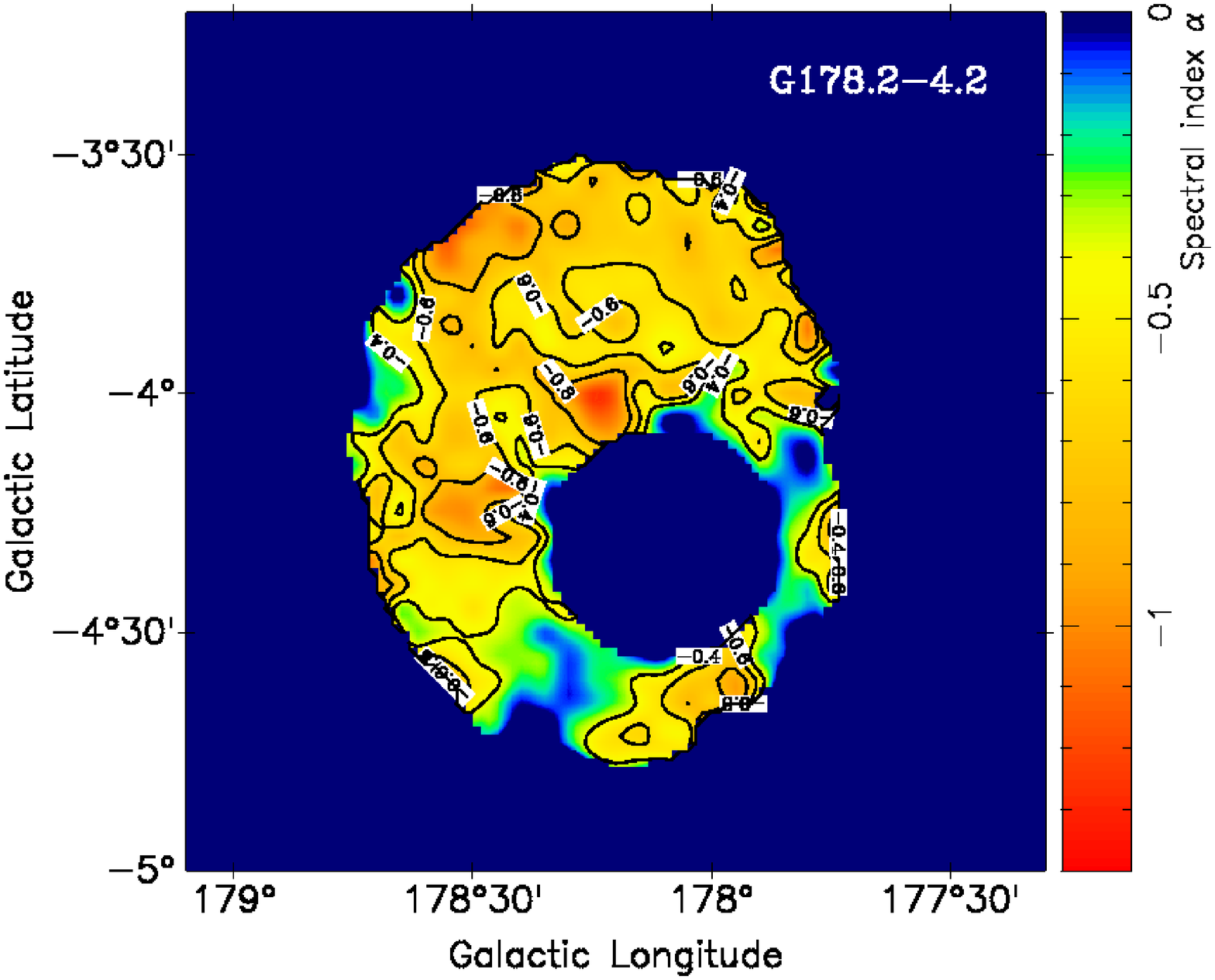}
\includegraphics[angle=-90, width=0.45\textwidth]{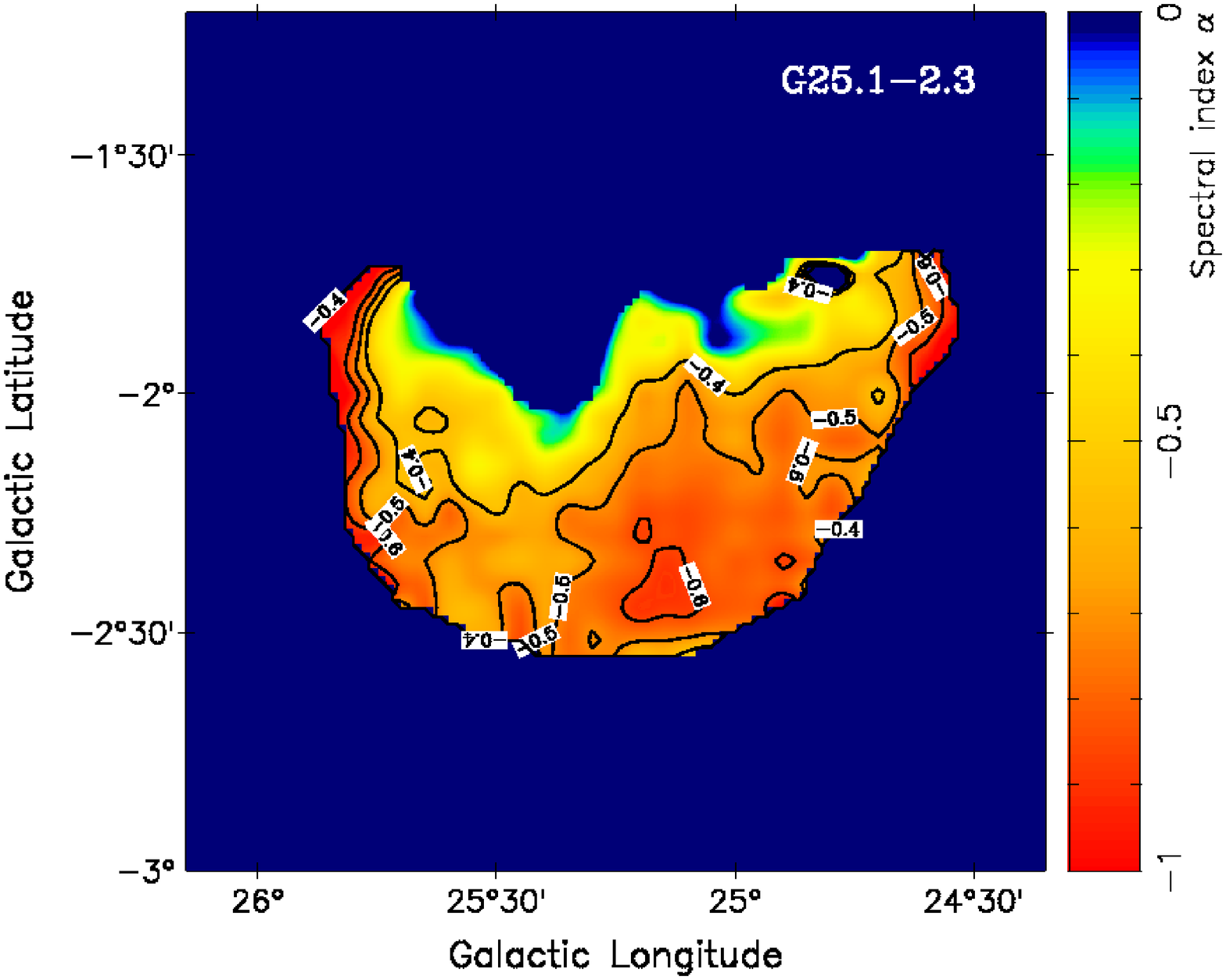}
\caption{Spectral index maps for G178.2$-$4.2 and G25.1$-$2.3.}
\label{spec_map}
\end{figure}

Spectral index maps for these two objects were derived from radio maps
at three wavelengths, as shown in Fig.~\ref{spec_map}. The spectral
indices vary in different areas. The spectral index of the northern
shell of G178.2$-$4.2 is about $\alpha \sim -0.6$, and that of the
southern shell of G25.1$-$2.3 is about $\alpha \sim -0.5$.
\begin{figure}
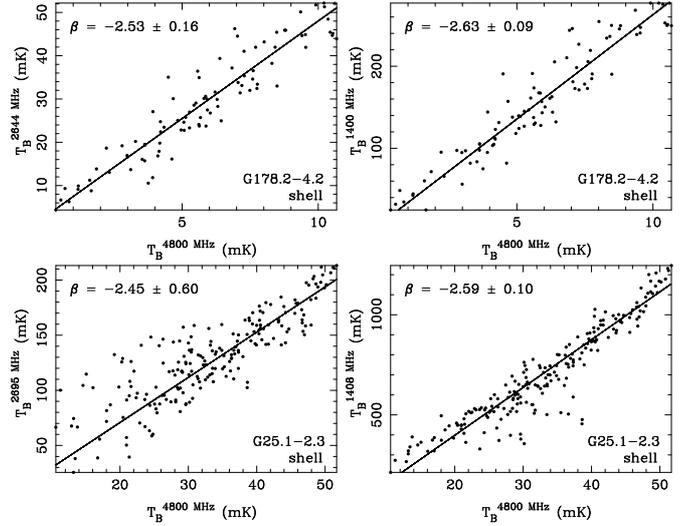

\includegraphics[angle=-90, width=0.235\textwidth]{G178_6_11_3min.ps}
\includegraphics[angle=-90, width=0.235\textwidth]{G178_6_21_3min.ps}\\
\includegraphics[angle=-90, width=0.235\textwidth]{G25_6_11.ps}
\includegraphics[angle=-90, width=0.235\textwidth]{G25_6_21.ps}
\caption{TT-plots for the shell of G178.2$-$4.2 outlined in
  Fig.~\ref{g178_maps} and the shell of G25.1$-$2.3 outlined in
  Fig.~\ref{g25_maps}, using the Urumqi $\lambda$6\ cm map and the
  Effelsberg $\lambda$11\ cm and $\lambda$21\ cm maps.}
\label{tt}
\end{figure}
These spectral indices were verified by TT-plots
\citep{Turtle62}. From the brightness temperatures $T_B$ at two
frequencies plotted against each other, the brightness spectral index,
$\beta$, is defined as $T_B = \nu^{\beta}$, so that $\alpha = \beta
+2$. As shown in Fig.~\ref{tt}, we got $\beta_{\rm 6cm-11cm} =
-2.53\pm0.16$ and $\beta_{\rm 6cm-21cm} = -2.63\pm0.09$ for the
northern shell of G178.2$-$4.2, and $\beta_{\rm 6cm-11cm} =
-2.45\pm0.60$ and $\beta_{\rm 6cm-21cm} = -2.59\pm0.10$ for the
southern shell of G25.1$-$2.3. These values all agree with the results
derived from the integrated flux densities and spectral index maps.

Using the total intensity and the radio spectral index, we calculated
the surface brightness of both objects at 1~GHz
\citep[e.g.][]{Kothes98}, obtaining $\Sigma_{\rm 1~GHz} = 1.505 \times
10^{-19} S_{\rm 1GHz} {\rm [Jy]} / {\square} {\rm [arcmin^2]} =
7.2\times10^{-23} {\rm Wm^{-2} Hz^{-1} sr^{-1}}$ for G178.2$-$4.2, and
$\Sigma_{\rm 1~GHz} = 5.0\times10^{-22} {\rm Wm^{-2} Hz^{-1} sr^{-1}}$
for the southern shell of G25.1$-$2.3. The surface brightness for
G178.2$-$4.2 is to date the second lowest for a Galactic SNR, and only
slightly above that of the lowest, $5.8\times10^{-23} {\rm Wm^{-2}
  Hz^{-1} sr^{-1}}$ reported for SNR G156.2+5.7 \citep{Reich92}.

Polarization observations of the G178.2$-$4.2 field are available at
$\lambda$6\ cm, $\lambda$11\ cm, and $\lambda$21\ cm.
The diffuse extended polarized emission at $\lambda$21\ cm (see the
{\it right bottom panel} in Fig.~\ref{g178_maps}) varies and has no
obvious structural relation to G178.2$-$4.2. We therefore conclude
that at $\lambda$21\ cm, most probably we see polarized diffuse
foreground or background Galactic emission and will not consider it in
the following discussion.

At $\lambda$11\ cm, the new sensitive Effelsberg observations of
G178.2$-$4.2 not only show significant polarized emission in the
northern shell but also two large patches of weak polarized regions
(see the {\it middle right panel} in Fig.~\ref{g178_maps}). The shell
structure is better resolved at $\lambda$11\ cm than at
$\lambda$6\ cm, and the detected polarized emission appears strong
along the shell ridge. The small polarization patch near $l =
177\degr50\arcmin$, $b = -4\degr10\arcmin$ has a very different
morphology when compared with the total-intensity map, and hence is
most likely not the emission from the SNR. The large polarization
patch between $178\degr40\arcmin > \ell > 178\degr10\arcmin$ and
$-3\degr50\arcmin > b > -4\degr20\arcmin$ has a polarization
percentage exceeding 100\%. The total-intensity radio emission from
the SNR is very weak and flocculent in this area, while the polarized
emission is significant and continuous, which indicates that this
polarized emission patch is likely also not physically related to the
SNR, otherwise it should also appear at $\lambda$6\ cm.  However,
neither of the $\lambda$11\ cm polarization patches has a counterpart
at $\lambda$6\ cm, which strongly suggests that these polarized
$\lambda$11\ cm features originate from unrelated Galactic emission within the
interstellar medium.  Polarization along the shell is clearly detected
at both frequencies.

At $\lambda$6\ cm, strong polarized emission is detected from the
northern shell of G178.2$-$4.2, while weak polarized emission is
spread over a large area both within and outside of the SNR. These
weak polarization patches are diffuse and extended and have a
brightness of a few mK $T_B$.  The lower and the central patches
inside the SNR are clearly not related to the total-intensity emission
and thus unlikely related to the SNR. Therefore, the polarized
emission reliably detected from G178.2$-$4.2 at both $\lambda$6\ cm
and $\lambda$11\ cm is of the shell region. The polarization angles of
the two frequencies differ by about 20$\degr$, corresponding to a
rotation measure $RM$ of about 36 + n $\times$ 350\ rad~m$^{-2}$ (n =
$\pm$1, $\pm$2 ...) by accounting for the n$\pi$-ambiguity. Because of
low $RM$ values in the anti-center region
\citep[e.g.][]{Spoelstra84,Sun08}, $n = 0$ is the most reasonable
choice, so that $RM$ = 36 \ rad~m$^{-2}$. This value is slightly
larger but comparable to $RM= 22.1\pm2.2$~rad\ m$^{-2}$ for the source
NVSS J052423+281232 at $\ell = 178\fdg09, b = -4\fdg31$ within the SNR
and $RM=11.8\pm17.1$~rad\ m$^{-2}$ for NVSS J052734+285134 at $\ell =
177\fdg94, b= -3\fdg37$ outside the SNR \citep{Taylor09} and the mean
$RM$ value of about 10~rad~m$^{-2}$ found for the nearby SNR S147
region \citep{Xiao08}.

Polarization observations of the G25.1$-$2.3 field are available at
$\lambda$6\ cm, and $\lambda$11\ cm.  Some weak polarization patches
have been detected at $\lambda$6\ cm and $\lambda$11\ cm (see
Fig.~\ref{g25_maps}, {\it right panels}). They have a very different
morphology to the total-intensity emission, and seem not to be
associated with the southern shell of G25.1$-$2.3. In general,
intrinsic polarized emission of a SNR should be detectable at a higher
frequency. This is not the case for G25.1$-$2.3. Comparing the
morphologies of the $\lambda$6\ cm and $\lambda$11\ cm polarization
map of G25.1$-$2.3, we conclude that the detected polarized radio
emission at both wavelengths is not associated with the SNR. This is
not unusual in this direction of the inner Galaxy, where the polarized
emission originates from diffuse emission in a complex environment
within the $\lambda$6\ cm polarization horizon of about 3~kpc
\citep[see][]{Sun11}.

\begin{figure}[hbt]
\includegraphics[angle=-90, width=0.44\textwidth]{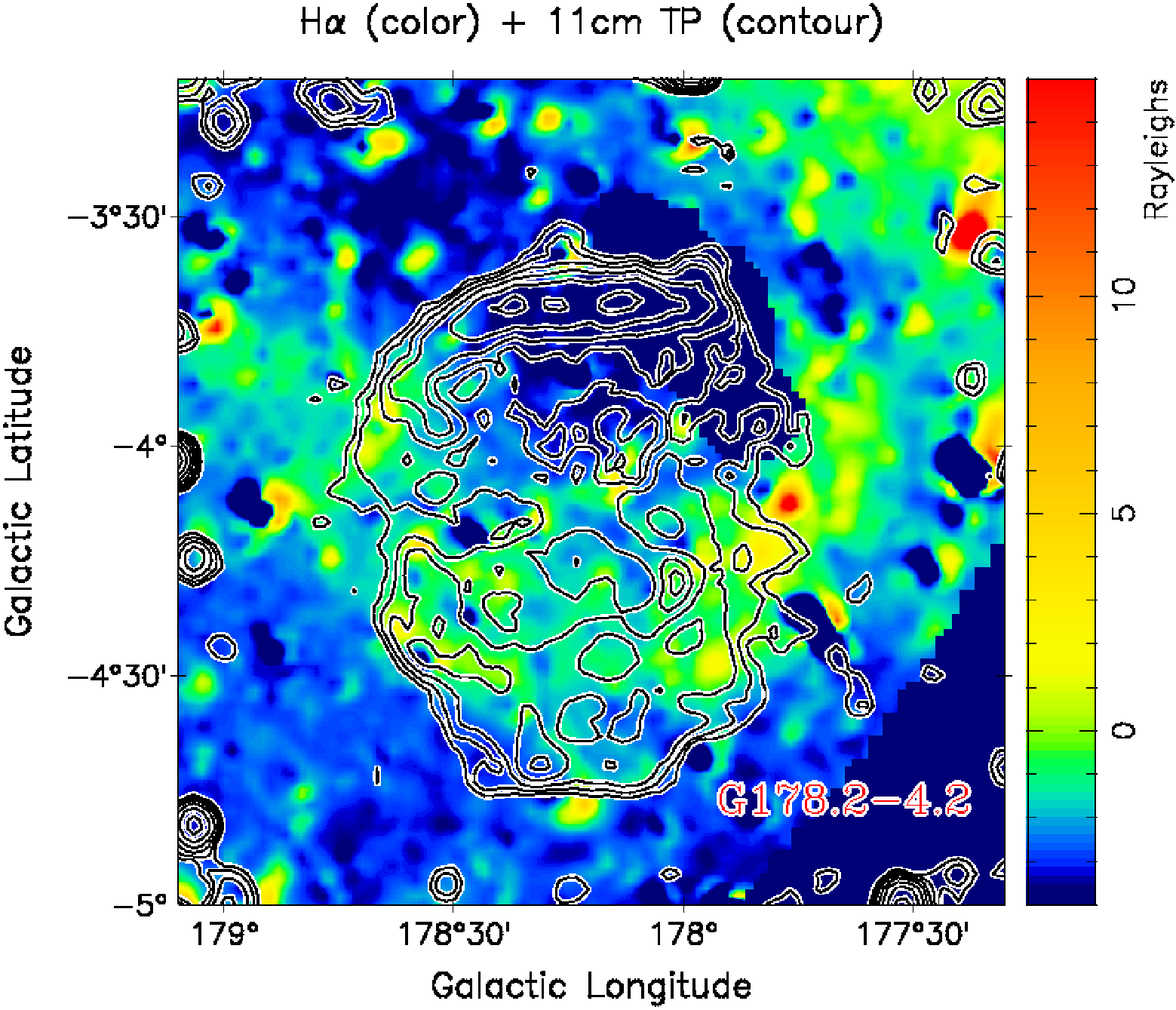}\\
\includegraphics[angle=-90, width=0.44\textwidth]{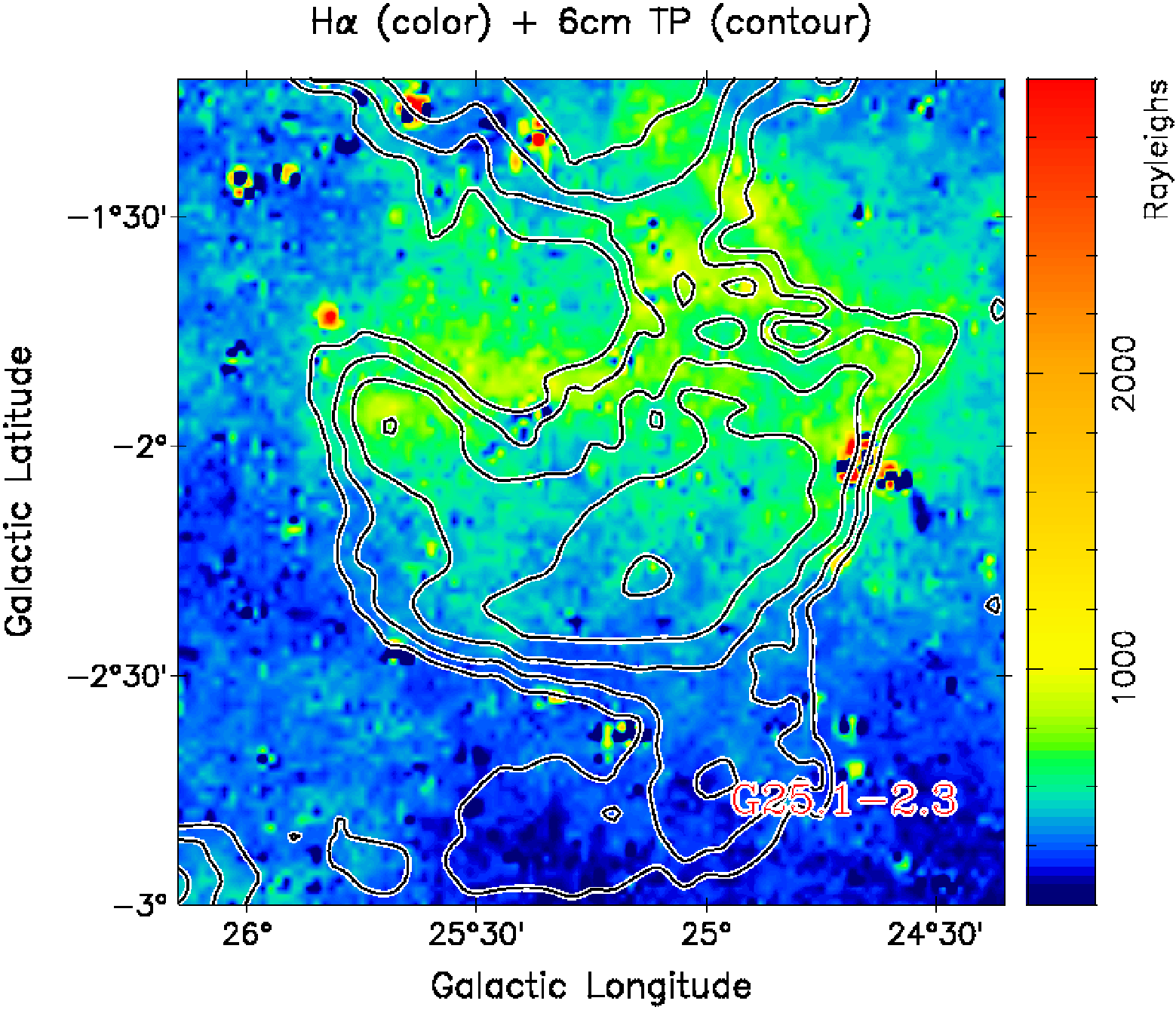}
\caption{Total-intensity contour maps of G178.2$-$4.2 at
  $\lambda$11\ cm and G25.1$-$2.3 at $\lambda$6\ cm with point-like
  sources subtracted overlaid onto VTSS H$\alpha$ image for
  G178.2$-$4.2 ({\it upper panel}) and SHASSA H$\alpha$ image for
  G25.1$-$2.3 ({\it lower panel}).}
\label{Halpha}
\end{figure}

\section{Signatures in other bands}

We have overlaid the radio maps of these two SNRs onto images from
other bands to see whether there is any spatial coincidence.

We superimposed the $\lambda$6\ cm total intensity contours of
G178.2$-$4.2 and G25.1$-$2.3 onto the images of the integrated CO
image \citep{Dame01}, the soft X-ray image in the 0.4-2.4~keV band and
the super-soft X-ray image in the 0.1-0.4~keV
band\footnote{http://www.xray.mpe.mpg.de/cgi$_{-}$bin/rosat/rosat$_{-}$survey}.
We could find no structural coincidence in these images with
G178.2$-$4.2 and G25.1$-$2.3.

Using ``The Virginia Tech Spectral-Line Survey
(VTSS)\footnote{http://www.phys.vt.edu/{\texttildelow}halpha/}'' and
``The Southern H-Alpha Sky Survey Atlas
(SHASSA)\footnote{http://amundsen.swarthmore.edu/}'' \citep{Gaustad01}
for arcmin-resolution digital images of interstellar H$\alpha$
emission, we have overlaid the total-intensity contour map of
G178.2$-$4.2 at $\lambda$6\ cm onto the VTSS H$\alpha$ image and that
of G25.1$-$2.3 at $\lambda$11\ cm onto the SHASSA H$\alpha$ image
(Fig.~\ref{Halpha}). In the area of G178.2$-$4.2, a very weak broad
band of H$\alpha$ emission runs across the SNR area and is considered
to be background or foreground emission. In the field of G25.1$-$2.3,
a strong H$\alpha$ emission patch is found to the north of the radio
shell. The H$\alpha$ emission appears to be weak within the shell
areas of both G178.2$-$4.2 and G25.1$-$2.3. Therefore, we consider
that the H$\alpha$ emissions present in these directions are probably
not physically associated with the SNRs.

\begin{figure}[hbt]
\includegraphics[angle=-90, width=0.4\textwidth]{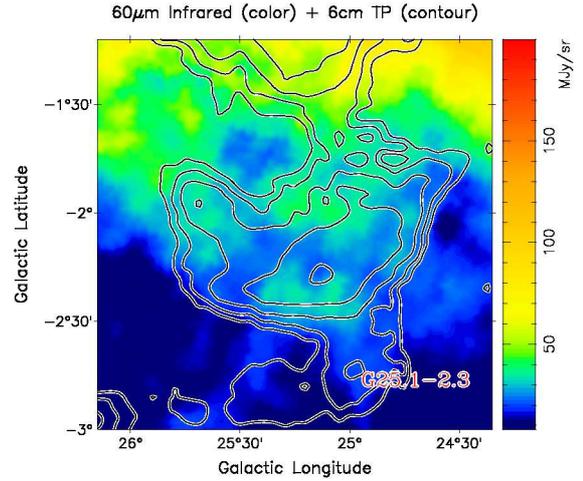}
\caption{Contour map of G25.1$-$2.3 at $\lambda$6\ cm overlaid onto
  the IRIS 60$\mu$m infrared image.}
\label{g25infra}
\end{figure}

\begin{figure}[hbt]
\centering
\includegraphics[angle=-90, width=0.32\textwidth]{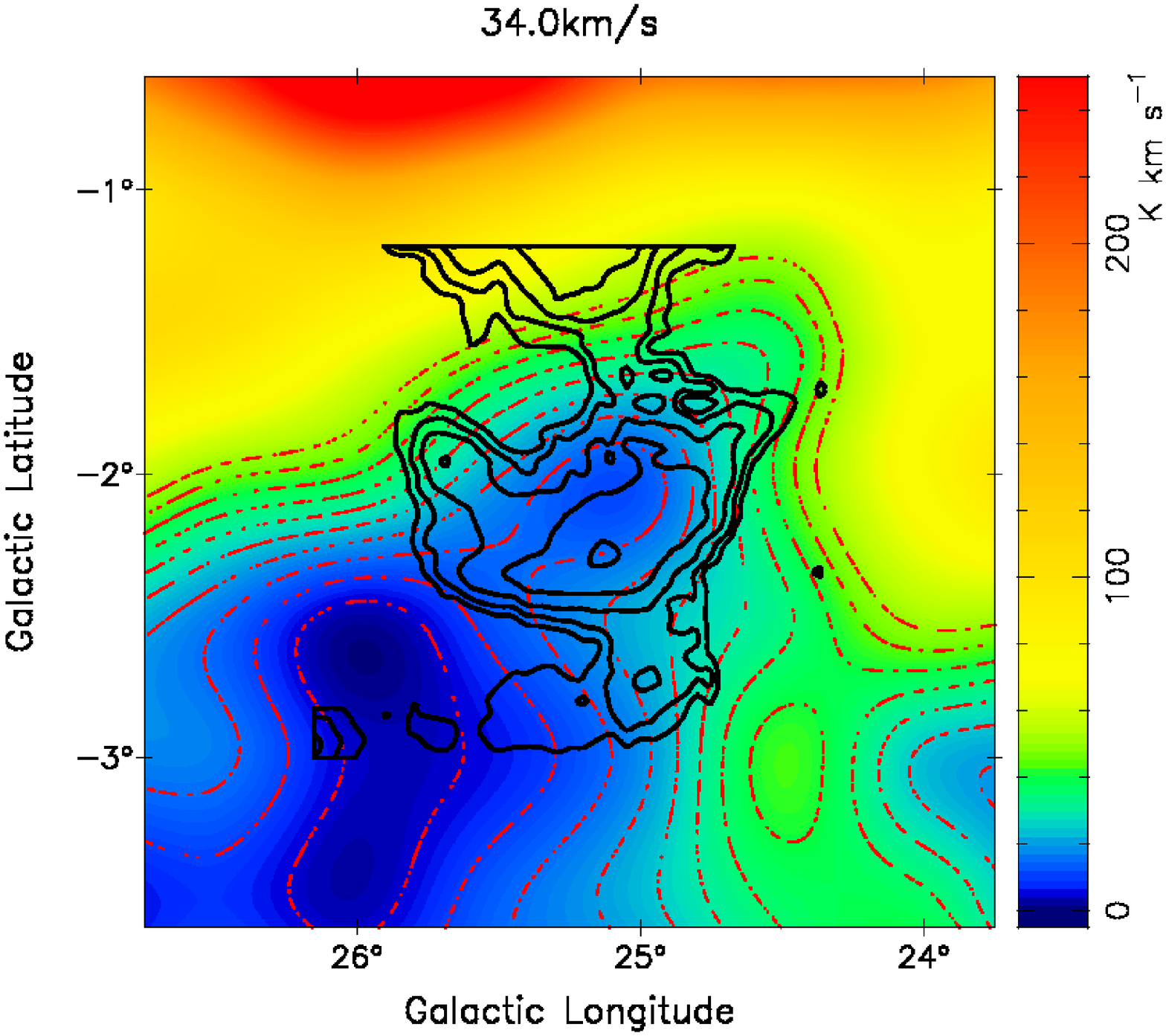}\\
\includegraphics[angle=-90, width=0.32\textwidth]{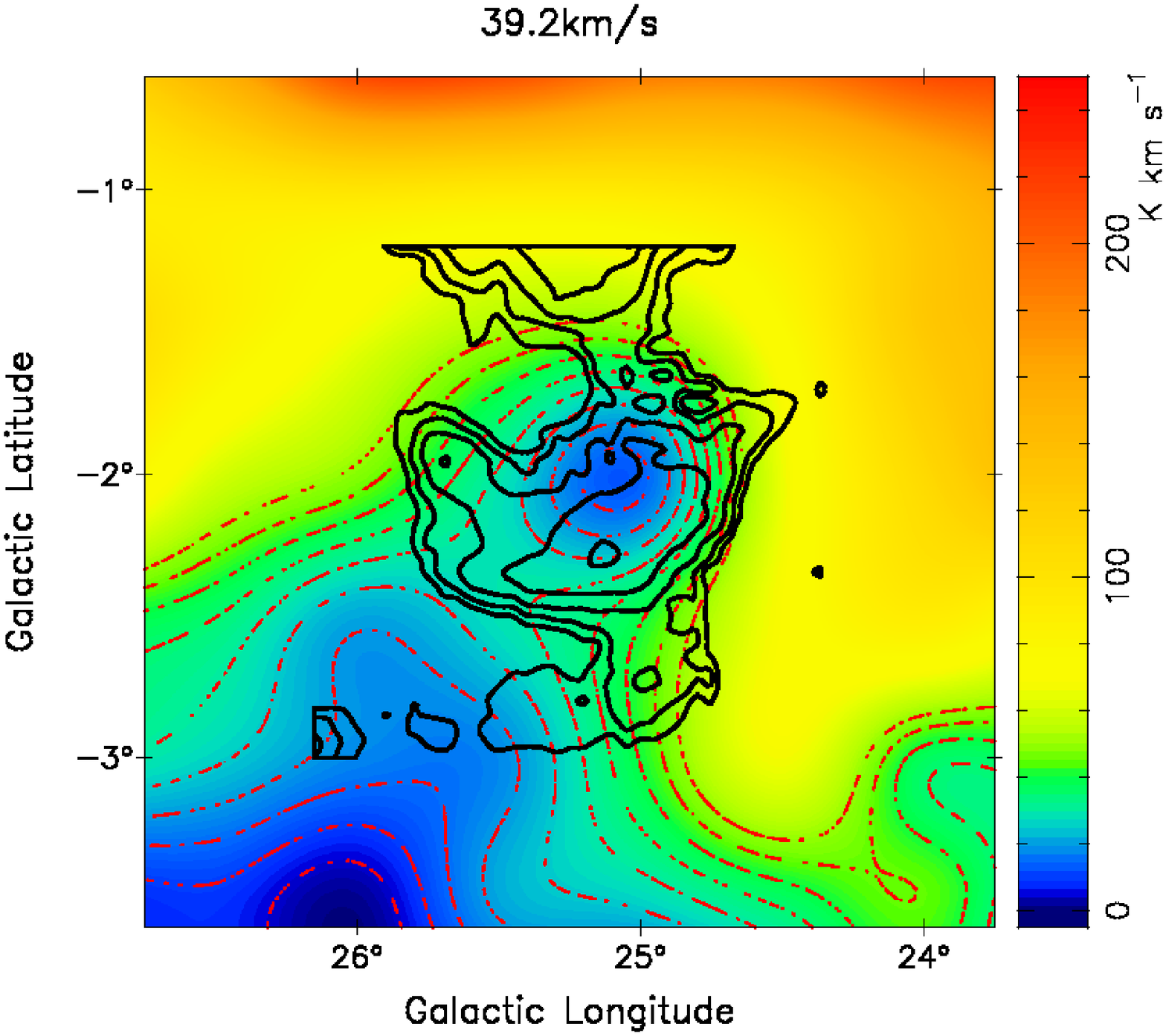}\\
\includegraphics[angle=-90, width=0.32\textwidth]{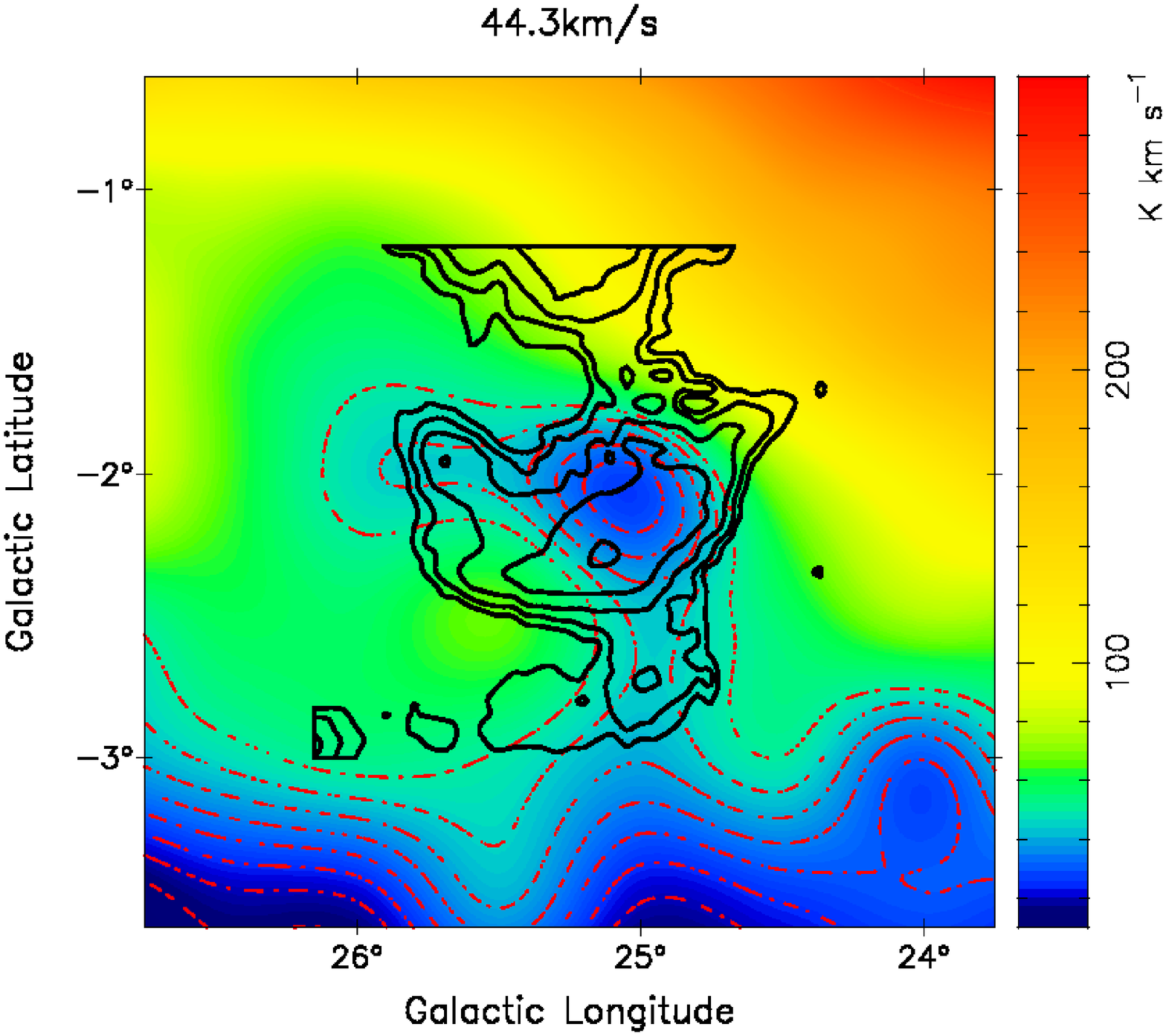}
\caption{\ion{H}{I} survey images with red dash-dot contour lines in
  the three velocity ranges 32.0 - 36.1, 37.1 - 41.2, and 42.2 - 46.4
  km s$^{-1}$ in the area of G25.1$-$2.3. The \ion{H}{I} cavity seems
  to have a morphological coincidence with the radio map of
  G25.1$-$2.3 (thick black contours).}
\label{HI_map}
\end{figure}

We compared the radio maps of G178.2$-$4.2 and G25.1$-$2.3 with the
IRIS\footnote{http://www.ias.fr/IRIS/} 60~$\mu$m infrared images
\citep{mml2005}. No coincidence was found for
G178.2$-$4.2. Fig.~\ref{g25infra} shows G25.1$-$2.3, where the
infrared patch positions have clear offsets to the radio
shell. Therefore, we conclude that both G178.2$-$4.2 and G25.1$-$2.3
have no corresponding infrared emission.

Very often SNRs show an associated \ion{H}{I} cavity. The
Leiden/Argentina/Bonn (LAB) \ion{H}{I} survey data with a velocity
resolution of 1.3~km/s \citep{Hartmann97, Kalberla05} was used to
search for \ion{H}{I} structures related to the SNRs in the velocity
range from $-$450 to 400~km/s, although the angular resolution of
36$\arcmin$ for the \ion{H}{I} data is coarse.
%
% G178.2$-$4.2
In the area of G178.2$-$4.2, there is a $3\degr \times 2\degr$
\ion{H}{I} shell centered at $\ell = 178\fdg2, b =-4\fdg1$, seen at a
central radial velocity of $-$3.2~km/s over a velocity range of
17.5~km/s \citep{Ehlerova05}. This is clearly larger than the SNR
extent, and its relation to the SNR cannot be settled.
%
% G25.1$-$2.3
%
In the area of G25.1$-$2.3, an \ion{H}{I} cavity is found in the
velocity range of 37.1 to 41.2\ km/s (see Fig.~\ref{HI_map}), which
has a very similar shape to G25.1$-$2.3 and hence might be related to
the SNR. The kinetic distance for the central velocity of 39.2\ km/s
is 2.9 or 12.4~kpc according to the Galactic rotation model by
\citet{Fich89}. These two values correspond to a span of 67~pc or
288~pc of the shell structure, respectively. Considering the physical
size of the SNR shell, the nearer distance is preferred.

For some SNRs, the empirical relation between the surface brightness
$\Sigma$ and diameter $D$ \citep[e.g.][]{Clark76} is the only tool to
estimate their distances, though with large uncertainties
\citep[e.g.][]{Green84, Green04}. We used the updated $\Sigma-D$
relation of \citet{Case98} to estimate the distances of both objects
we newly found. For G178.2$-$4.2, we found the corresponding diameter
of that surface brightness is 197~pc, so that its distance is 9.4~kpc,
which places the object very outside the Galaxy and seems not
possible. For G25.1$-$2.3, we found the diameter for the shell is
72~pc and the distance is 3.1~kpc, consistent with those derived from
the HI data. Note, however, that the uncertainties of these estimates
could be as large as 40\%.

Several pulsars are known within the field of G25.1$-$2.3
(Fig.~\ref{g25_maps}).  They have distances of 4 $-$ 6\ kpc according
to the pulsar dispersion measure and the electron distribution model
of the Galaxy, NE2001 \citep{Cordes02}. It is not possible to
associate any of these pulsars with the SNR unambiguously.

\section{Summary}

We have found two shell-like objects, G178.2$-$4.2 and G25.1$-$2.3,
from the radio map of the Sino-German $\lambda$6\ cm polarization
survey of the Galactic plane. In addition, using the Effelsberg
$\lambda$11\ cm and $\lambda$21\ cm continuum and polarization maps, a
shell-type morphology is confirmed for both. Their radio spectra are
characteristic of non-thermal emission with spectral indices for the
shells of $\alpha \sim -0.6$ for G178.2$-$4.2 and $\alpha \sim -0.5$
for G25.1$-$2.3. Such values are typical for SNRs.  An ordered
magnetic field runs along the northern shell of G178.2$-$4.2. An
\ion{H}{I} cavity, likely at a distance of about 2.9~kpc, is probably
associated with the SNR G25.1$-$2.3. This would imply a diameter of up
to 67~pc in size.

\begin{acknowledgements}
We like to thank the anonymous referee for helpful comments. The Sino-German
$\lambda$6\ cm polarization survey was carried out with a receiver
system constructed by Mr. Otmar Lochner at MPIfR mounted at the
Nanshan 25-m telescope of the Urumqi Observatory of NAOC. The MPG and
the NAOC/CAS supported the construction of the receiving system by
special funds.  We thank Mr. Maozheng Chen and Mr. Jun Ma for
qualified maintenance of the receiving system for many years. The new
$\lambda$11\ cm map of G178.2$-$4.2 is based on observations with the
100-m telescope of the MPIfR at Effelsberg.
The Chinese authors are supported by the National Natural Science
foundation of China (10773016, 10821061, and 10833003), the National
Key Basic Research Science Foundation of China (2007CB815403), and the
Partner group of the MPIfR at NAOC in the frame of the exchange
program between MPG and CAS for many bilateral visits.
XYG thanks the joint doctoral training plan between CAS and MPG and
the financial support from CAS and MPIfR. XHS thanks the MPG and Prof.
Michael Kramer for financial support during his stay at the MPIfR.
We used the H$\alpha$ images for G178.2$-$4.2 from the Virginia Tech
Spectral-Line Survey (VTSS), which is supported by the National
Science Foundation. We thank Prof. Ernst F\"{u}rst for critically
reading the manuscript.
\end{acknowledgements}

\bibliographystyle{aa}
\bibliography{bbfile}

\end{document}